\documentclass[journal,onecolumn]{IEEEtran}
\usepackage{amsthm}
\usepackage{graphicx}
\usepackage{algorithm}
\usepackage{algorithmic}
\newtheorem*{Definition}{Definition}

\usepackage{caption}
\usepackage{subfigure}
\usepackage{url}
\usepackage[utf8x]{inputenc} 
\usepackage{cite}
\usepackage{bbding}
\usepackage{pifont}
\usepackage{pdflscape}
\usepackage{longtable}
\usepackage[english]{babel}
\usepackage{array}
\newcolumntype{L}[1]{>{\raggedright\let\newline\\\arraybackslash\hspace{0pt}}m{#1}}
\newcolumntype{C}[1]{>{\centering\let\newline\\\arraybackslash\hspace{0pt}}m{#1}}
\newcolumntype{R}[1]{>{\raggedleft\let\newline\\\arraybackslash\hspace{0pt}}m{#1}}
\usepackage{wrapfig}
\usepackage{tikz}
\usepackage{amsmath}
\usepackage{amsfonts}
\usepackage{amssymb}
\usepackage{mathrsfs}
\usepackage{ragged2e}
\usepackage{verbatim}
\usepackage{rotating,booktabs}
\def \Z {{\Bbb Z}}
\def \R {{\Bbb R}}

\def \F {{\Bbb F}}

\newtheorem{theorem}{Theorem}

\newtheorem{example}[theorem]{Example}
\usepackage[english]{babel}
\usepackage{fancyhdr}
\usepackage{lastpage}
 \usepackage{wrapfig}
\begin{document}
\vspace{5cm}
\title{The Art of DNA Strings: Sixteen Years of DNA Coding Theory}
\author{
\IEEEauthorblockN{Dixita Limbachiya, Bansari Rao and Manish K. Gupta}
\IEEEauthorblockA{\\
Dhirubhai Ambani Institute of Information and Communication Technology\\
Gandhinagar, Gujarat, 382007 India\\
Email:dlimbachiya@acm.org, bansari\_s\_rao@daiict.ac.in and mankg@computer.org\\
}
}
 
\pagenumbering{arabic}
\maketitle
\IEEEpeerreviewmaketitle

\begin{abstract}
The idea of computing with DNA was given by Tom Head in 1987, however in 1994 in a seminal paper, the actual successful experiment for DNA computing was performed by Adleman. The heart of the DNA computing is the DNA hybridization, however, it is also the source of 
errors. Thus the success of the DNA computing depends on the error control techniques. The classical coding theory techniques have provided foundation for the current information and communication technology (ICT). Thus it is natural to expect that coding theory will be the foundational subject for the DNA computing paradigm. For the successful experiments with DNA computing usually we design DNA strings which are sufficiently dissimilar. This leads to the construction of a large set of DNA strings which satisfy certain combinatorial and thermodynamic constraints. Over the last 16 years, many approaches such as combinatorial, algebraic, computational have been used to construct such DNA strings. In this work, we survey this interesting area of DNA coding theory by providing key ideas of the area and current known results. 
\end{abstract}

\section{Introduction}

\begin{wrapfigure}{r}{0.25\textwidth}
  \begin{center}
    \includegraphics[width=0.19\textwidth]
    {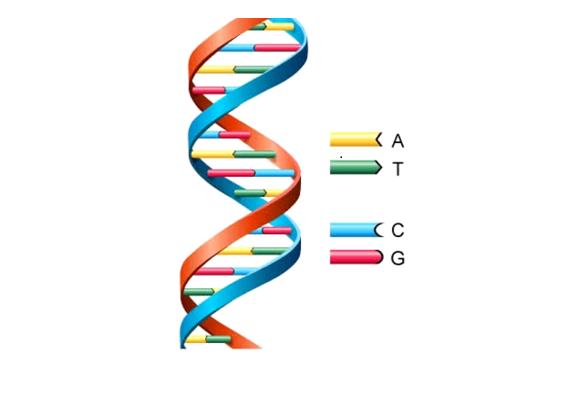}
  \end{center}
  \caption{\small{DNA structure with its four nucleotides $A$-Adenine, $G$-Guanine, $C$-Cytosine and $T$-Thymine. These are the basic building blocks of DNA which are held by Hydrogen bonds in a double helical manner. Each DNA base is paired with the complementary bases such that the yellow band is $A$ which is connected to its complementary base $T$ (green band) while $C$ (blue band) is paired with red $G$ (red band)}}
\end{wrapfigure}

Information and Communication Technology (ICT) has come a long way in the last $60$ years. We have now connected world of devices talking to each other or performing the computation. This is possible due to the advancement of coding and information theory. In $1994$, two new directions in ICT emerged viz. quantum computing and DNA computing. The heart of any computing or communications is coding theory. Thus two new directions in coding theory also emerged such as quantum coding theory and DNA coding theory. This paper focuses on the later area giving a comprehensive overview of techniques and challenges of DNA coding theory from last 16 years. 
In $1994$, L.Adleman performed the computation using DNA strands to solve an instance of the Hamiltonian path problem giving birth to DNA computing \cite{adleman1994molecular,maley1998dna,calude2000computing,garzon1997new}. DNA is a double strand built by the pairing the four basic building units A-(Adenine), C-(Cytosine), G-(Guanine), T-(Thymine) which are called nucleotides. The DNA strand is held by the important feature called complementary base pairing which connects the Watson Crick complementary bases with each other denoted by $A^C = T$ and $G^C = C$. The backbone of DNA is an alternating chains of sugar and phosphate. DNA is an ideal source of computing due to its stable, dense and its self replicating property  \cite{deaton1998reliability}. After the successful experiment by Adleman, area of DNA computing \cite{rozenberg1999dna} flourished into different directions like DNA tile assembly \cite{winfree1998design} \cite{winfree1998algorithmic}, building of DNA nano-structures \cite{seeman1998dna} \cite{rothemund2006folding}, studying error correcting properties in DNA sequences \cite{arita2004writing} \cite{d2003exordium} and DNA based data storage system \cite{limbachiya2015natural}. But it was only with the identification of mathematical properties in DNA sequences \cite{marathe2001combinatorial} \cite{hussini2002coding}, it inspired many researchers to explore this new amalgamation of biology and coding theory  \cite{gupta2006quest}. DNA computing \cite{watada2008dna} promises to solve many NP-complete problems like Hamilton path problem, satisfiability problem (SAT) \cite{wang2008solving} and many others with massive parallelization \cite{darehmiraki2010semi}. To perform computation using DNA strands, a specific set of DNA sequences are required with particular properties. Such set of DNA strands satisfying various constraints \cite{hartemink1999automated,li2008optimization} play an important role in biomolecular computation \cite{penchovsky2003dna} \cite{baum1999dna}. 

\begin{figure*}[ht]
\centering
\includegraphics[scale=0.35]{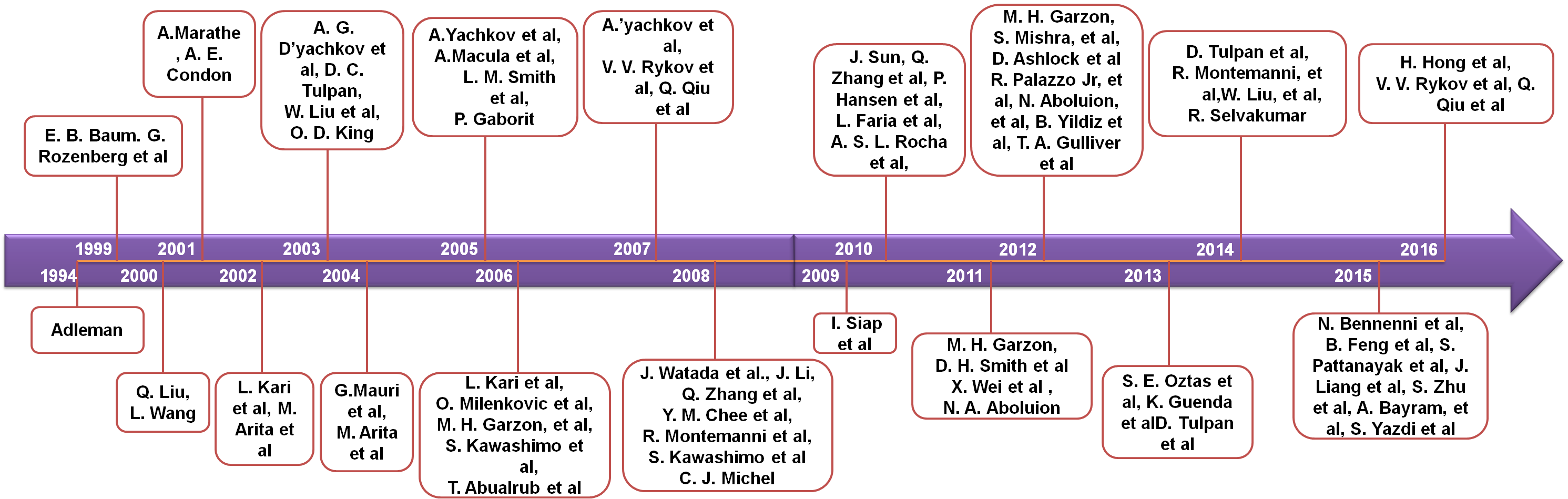}
\caption{DNA codes Time line: The state of art of DNA codes is showcased from $1994$ to $2016$. Different work mentioned with authors in chronological order.}
\label{timeline}
\end{figure*}

The aim of this manuscript is to elucidate the current state of art of DNA codes as shown in Figure \ref{timeline}, DNA codes constraints and especially the new methods contributing to construct DNA codes from algebraic coding. It summarizes the research on DNA codes which may help the researcher to get a broad picture of the area. Also, it gives brief overview on the applications of DNA codes.

The Paper is organized as follows. Section II and III discusses the DNA code design problem and constraints on DNA codes. Section IV describes various approaches for the construction of DNA codes. Section V gives a brief description on software tools for DNA sequence design. Section VI includes description on bounds of DNA codes. Section VII gives a brief note on applications of DNA codes. Section VIII shows results on DNA codes for $0\leq n\leq36$ and $0\leq d\leq20$. Section IX concludes the paper with general remarks.

\section{DNA Code Design Problem}

To perform the DNA computation, DNA strands react with each other by Watson Crick base pairing and form perfect match. But in some situation, DNA strands may not form perfect base pairing and react in undesirable manner. One situation is formation of secondary structure in which first half strand of the DNA strand forms complementary with its own other half forming the hair pin like structure. This kind of structure interrupt the desired computation. Such secondary structures have to be avoided by designing the DNA strands carefully. Also DNA strands may bind to another DNA strands forming the complementary base pairing with few base pairs creating an error. One way to avoid this is to ensure that every two DNA strands differ in more than $d$ locations where $d$ depends on the application of DNA computation. This property can be obtained by defining different distance like the Hamming distance, Lee distance, Edit distance, Deletion distance etc. between two DNA strands. 

DNA codes design problem \cite{d2005dna,milenkovic2006design,garzon2006search} is to develop a set of the DNA codewords of length $n$ over DNA alphabets $\Sigma_{DNA} = \{A, T, G, C\}$ with predefined distance $d$ \cite{dinu2006low} \cite{d2007dna} and satisfying maximum set of constraints \cite{sager2006designing}. The main objective is to find the largest possible set of codewords $M$ of length $n$ over alphabet $\Sigma_{DNA} = \{A, C, G, T\}$ feasible with respect to set of constraints \cite{sun2010bounds} with distance $d$ that enable the construction of better error correcting codes \cite{debata2012coding,ashlock2012synthesis,faria2012genome}. 

\begin{Definition}[DNA code]
A DNA code $\mathscr{C}_{DNA}(n,M,d) \subset \Sigma_{DNA} =\{A,T,G,C\}$ with each DNA codeword of length $n$ and size $M$ and minimum distance $d$. Here $A$ denotes Adenine, $T$ denotes Thymine, $G$ denotes Guanine and $C$ denotes Cytosine as the nucleotides in DNA.
\end{Definition}

\section{Constraints on DNA Codes}

There are different types of constraints that DNA codes must satisfy. Three categories in which these constraints can be classified are as follows:
\begin{enumerate}
\item Combinatorial constraints
\item Thermodynamic constraints
\item Application Oriented constraints
\end{enumerate}
\begin{figure*}[htbp]
\centering
\includegraphics[scale=0.35]{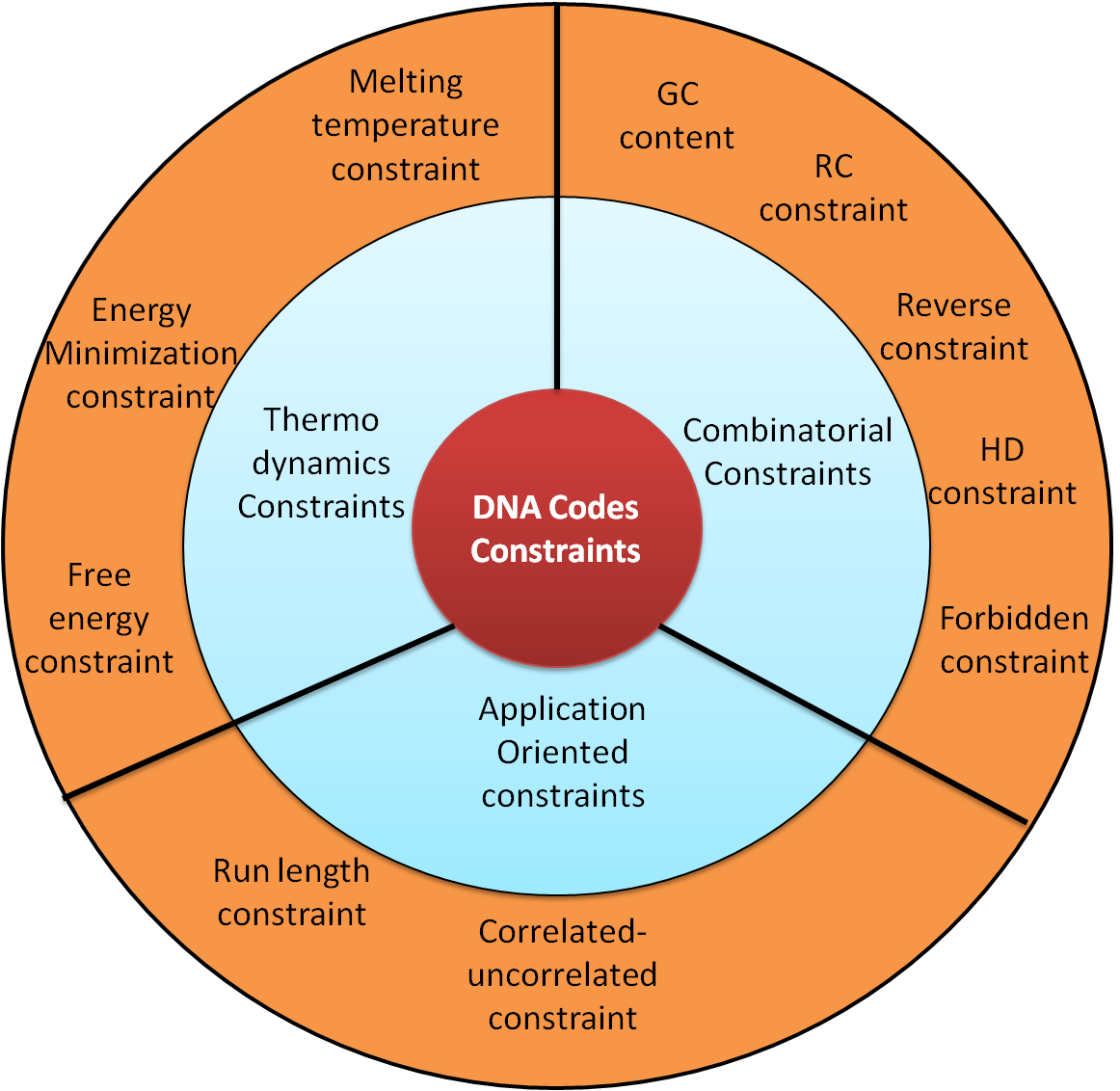}
\caption{DNA codes constraints classified as three categories combinatorial, thermodynamics and application oriented constraints.}
\end{figure*}

The constraints which should be followed by DNA codes are listed below : 
\begin{enumerate}

\item \textit{Hamming distance constraint $(n,d,w)$} -  

The Hamming distance constraint can be defined as  $d_{H}(\textbf{x}_{\textbf{DNA}}$, $\textbf{y}_{\textbf{DNA}})$ $\geq d $ $ \forall $ $\textbf{x}_{\textbf{DNA}}$, $\textbf{y}_{\textbf{DNA}} \in  \mathscr{C}_{DNA}$ for some Hamming distance $d$ \cite{marathe2001combinatorial}. A  set of codewords with length $n$, size $M$ and minimum Hamming distance $d$ satisfying Hamming constraint is denoted by $\mathscr{C}_{DNA}(n,M,d)$. Hamming distance is calculated by the total number of places at which two DNA codewords differ. For $\mathscr{C}_{DNA}(n,M,d)$, the minimum of the distances is considered as the Hamming distance. $A_q(n,d)$ denotes the maximum size of a code with codewords of length $n$ and distance $d$ over alphabet size $q$, in case of DNA codewords $q=4$. 

\begin{example}
Let $\textbf{x}_{\textbf{DNA}} = ATGACT$ and $\textbf{y}_{\textbf{DNA}} = ACTAGC$, then $d_{H}(\textbf{x}_{\textbf{DNA}}$, $\textbf{y}_{\textbf{DNA}})$ = $4$ where $\textbf{x}_{\textbf{DNA}}, \textbf{y}_{\textbf{DNA}} \in \mathscr{C}_{DNA}$ following the Hamming distance constraint with $d_{H}(\textbf{x}_{\textbf{DNA}}$, $\textbf{y}_{\textbf{DNA}}) \geq 3 $. 
\end{example}

\item \textit{Reverse constraint$(n,d)$} - The reverse constraint is $H_{DNA}(\textbf{x}_{\textbf{DNA}}^{\textbf{R}},\textbf{y}_{\textbf{DNA}}) \geq d $ $   \forall $  $ \textbf{x}_{\textbf{DNA}}$, $\textbf{y}_{\textbf{DNA}} \in  \mathscr{C}_{DNA}$. A code satisfying reverse constraint is called reverse code. $A^{R}_q(n,d)$ denotes the maximum size of a reverse code with length $n$ and minimum Hamming distance $d$.

\begin{example}
Let $\textbf{x}_{\textbf{DNA}}  = ATGACT$, $\textbf{x}_{\textbf{DNA}}^{\textbf{R}} = TCAGTA$ and $ \textbf{y}_{\textbf{DNA}} = ATACAT$.  For $n=6$ and $d=3$, $H_{DNA}(\textbf{x}_{\textbf{DNA}}^\textbf{R},\textbf{y}_{\textbf{DNA}}) \geq 3 $. $\textbf{x}_{\textbf{DNA}}$, $\textbf{y}_{\textbf{DNA}}$ are reverse code.
\end{example}

\item \textit{Reverse Complement constraint $(n,d)$ } - This RC-constraint is $H_{DNA}(\textbf{x}_{\textbf{DNA}}^{\textbf{R}},\textbf{y}_{\textbf{DNA}}^{\textbf{C}}) \geq d $ $ \forall$ $\textbf{x}_{\textbf{DNA}}$, $\textbf{y}_{\textbf{DNA}}$ $\in \mathscr{C}_{DNA}$. DNA code satisfying RC-constraint is called a reverse complement code. $A^{RC}_q(n,d)$ denotes the maximum size of a reverse complement code with length $n$ and minimum Hamming distance $d$ \cite{marathe2001combinatorial}.

\begin{example}
Let $\textbf{x}_{\textbf{DNA}}  = ATGACT$, $\textbf{x}_{\textbf{DNA}}^{\textbf{R}} = TCAGTA$ and $\textbf{y}_{\textbf{DNA}}=GTACAC$, $\textbf{y}_{\textbf{DNA}}^{\textbf{C}} = CATGTG$. For $n=6$ and $d=3$, $H_{DNA}(\textbf{x}_{\textbf{DNA}}^{\textbf{R}},\textbf{y}_{\textbf{DNA}}^{\textbf{C}})= 4 \geq 3 $. $\textbf{x}_{\textbf{DNA}}$, $\textbf{y}_{\textbf{DNA}}$ are reverse complement code.
\end{example}

\item \textit{$GC$-content constraint $(n,d,w)$} - The set of codewords with length $n$, distance $d$ and GC weight $w$ ,where $w$ is total number of $G$s and $C$s present in the DNA strand viz. $w_{\textbf{x}_{DNA}} = \vert \{x_i:\textbf{x}_{\textbf{DNA}}=(x_i)$ ,  $x_i\in\{C,G\}\}\vert$ \cite{chee2008improved,smith2011linear,tulpan2014thermodynamic}.  Generally 
$w = \lfloor{n/2}\rfloor$. 

\begin{example}
Let $\textbf{x}_{\textbf{DNA}} = ATTGCT$ then $\textbf{x}_{\textbf{DNA}}  \notin \mathscr{C}_{DNA}$ for $n=6$ and $w=3$.
\end{example}

\item \textit{Melting temperature constraint $(n,d,w)$}- Melting temperature $T_m$ is a temperature at which half of the DNA strands are hybridized and half are not. Melting is opposite of hybridization in which two strands get separated. It is advantageous to have the codewords with the similar melting temperatures as it will enable the hybridization of multiple DNA strands simultaneously. Hence we select the codewords with similar melting temperature calculated by Nussinov's algorithm \cite{milenkovic2006design}. For each $\textbf{x}_{\textbf{DNA}}  \in  \mathscr{C}_{DNA}$ have identical melting temperature\cite{sager2006designing} .

\item \textit{ Thermodynamic constraint $(n,d,w)$} - For the DNA stability, it is necessary  for all the DNA codes should have comparable free energy  $\Delta G^{\circ}$ \cite{bishop2007free,d2005weighted,zhang2011evaluating} above some threshold \cite{tulpan2005thermodynamically,zhang2010dna}. As DNA with minimum free energy is more stable and hence we consider only those DNA codewords in a DNA code that have approximately comparable free energy in the set of DNA codewords. Free Energy $\Delta G^{\circ}$ for the given DNA codewords  $\textbf{x}_{\textbf{DNA}}, \textbf{y}_{\textbf{DNA}}  \in \mathscr{C}_{DNA}$ can be obtained by the equation, \\
\begin{equation*}
    |\Delta G^{\circ}(\textbf{x}_{\textbf{DNA}}) - \Delta G^{\circ}(\textbf{y}_{\textbf{DNA}}) | \leq \delta
\end{equation*} where $\delta > 0$ is a constant.

\item \textit{ Uncorrelated-correlated constraint $(n,d,w)$} - A codeword $\in$ $\mathscr{C}_{DNA}$ if shifted by $x$ units where $x \leq$ $n$ should not match with any of the other codeword $\in$ $\mathscr{C}_{DNA}$ \cite{yazdi2015rewritable}.
\\
\begin{example}
Let $\textbf{X}_{\textbf{DNA}}= CATCATC$ and $\textbf{Y}_{\textbf{DNA}}= ATCATCGG$.
$X \circ Y = 0100100$, as depicted below.\\ \\
\begin{tabular}{l l l l l l l l l l l l l l l l l l}
X =& C & A & T & C & A & T & C & & & & & & & & & &  \\
Y =& A & T & C & A & T & C & G & G & & & & & & & & & 0\\
Y =& & A & T & C & A & T & C & G & G & & & & & & & &1\\
Y =& & & A & T & C & A & T & C & G & G & & & & & & &0\\
Y =& & & & A & T & C & A & T & C & G & G & & & & & &0\\
Y =& & & & & A & T & C & A & T & C & G & G & & & & &1\\
Y =& & & & & & A & T & C & A & T & C & G & G & & & &0\\
Y =& & & & & & & A & T & C & A & T & C & G & G & & &0\\ \\
\end{tabular}
\end{example}
\end{enumerate}

The constraints (1) to (3) are used to avoid undesirable hybridization between different DNA strands and (4) to (6) are the DNA constraints which ensures that all the codewords have similar thermodynamic characteristics to perform uniform computation.

\section{Various Approaches for the DNA Codes Constructions}
There are different approaches \cite{tulpan2006effective} for the construction of the DNA codes with finite length $n$, defined distance $d$ and set of constraints with respect to application. The set of constraints essential for the DNA codes is subject to the application.  There exist algorithmic, theoretic and software simulation method \cite{feldkamp2003software} \cite{feldkamp2000dna} \cite{
feldkamp2001dnasequencegenerator} approaches to design the DNA codes. Optimality of the DNA code can be obtained by construction of the DNA codes in a way that every codeword in the set follows maximum number of constraints for a large value of $n$ and large minimum distance $d$ with minimum errors in DNA computation \cite{gray1996reducing}. 

\begin{figure*}[ht]
\centering
\includegraphics[scale=0.5]{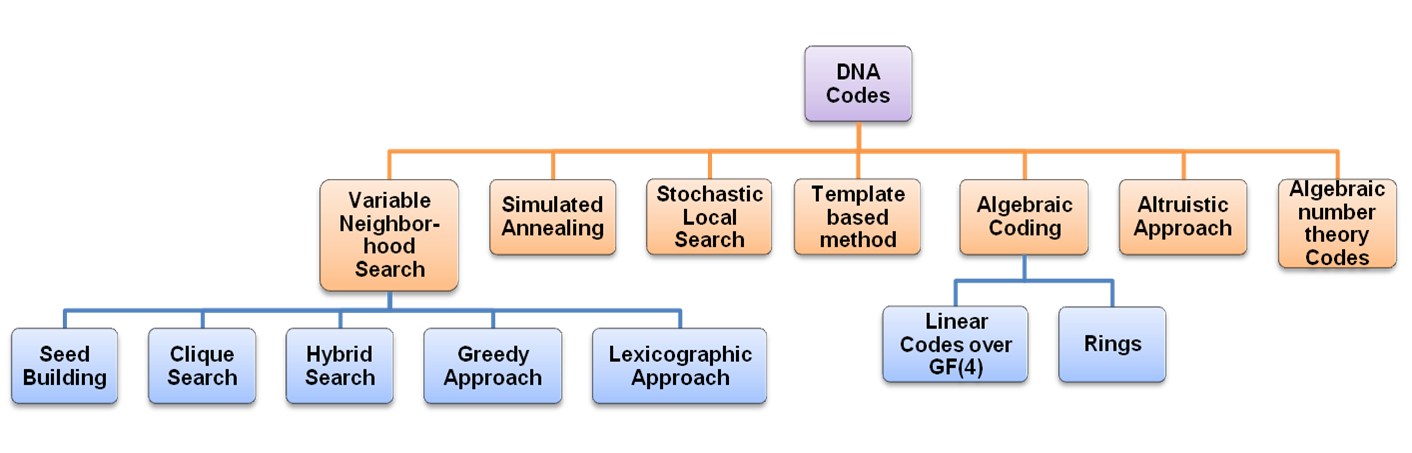}
\caption{DNA codes construction Methods are classified as Search algorithms, Algebraic Coding methods and Software simulations. Search algorithms include Variable neighborhood search, Simulated Annealing, Stochastic Local Search. Template Based method uses a template to design a DNA codeword. Algebraic coding method is construction of DNA codes by using Algebraic structures like Fields and Rings. Altruistic Approach is construction of DNA codewords in altruistic way in this work. }
\end{figure*}

\begin{enumerate}
 
\item \textbf{Variable Neighborhood Search Approach:} This approach employs different local search algorithms to search the DNA codes \cite{hansen2010variable,kawashimo2006dna,montemanni2008construction,kawashimo2008dynamic}. Some of the search algorithms are described here.

\begin{enumerate}
\item\textit{Seed Building(SB) :} Seed Building (SB) algorithm examines all the possible codewords randomly with respect to seed codeword where seed codewords are the initial set of codewords with the required constraint. For more details on the algorithm, \cite{montemanni2014three} can be referred. The drawback of the approach is that it is inefficient for larger values of $n$ because of the computational and time complexity invested for the development of feasible set of codewords.

\begin{example}
Consider the $n=3, d=1$ and $GC$-content $w=2$ and initial seed be codewords with $d=1$ and $w =2$ are $AGG$ and $AGC$ then the resulting codewords at first iteration with HD and GC-content constraints are  $GGC,GCA, GTC, CCG, CGC, CTC, TGG$. Note that $ACC, GCT,CAG$ are also codewords with $GC$-content $2$ but at distance $d=3$ so will be added in the next iteration for seed $d=3$.
\end{example}

\item \textit{Clique Search :} In this method a random subset of the codewords of a given code is removed, leaving a partial
code. All the codewords are generated and those compatible with the ones left in the code are identified and a graph is built where the identified codewords are nodes, and an edge exists between compatible codewords. 

\begin{example}
Let $n=4, d=3$ and $w=2$ then the partial code be $CTTC, CGAA, TGGT, GTGA$ with HD, RC and GC constraints. The clique search will result in codewords $CACT, GCTT, AGTG$ and $AAGC$. 

\end{example}

\item \textit{Hybrid Search:} This method combines two approaches of seed building and clique search. It uses seed building to generate the partial code and clique search to search for the best codewords \cite{qiu2007hybrid}.

\item \textit{Greedy Approach:} These types of algorithms removes the worst DNA code at each iteration from a set of codes every time the algorithm is iterated. But the problem with this approach is that it doesn't always produce the best results. In the process of removing the worst code at each stage it may remove a potentially good code at earlier stages and may not remove bad code at the later stages. So it doesn't always produce the best optimal result \cite{Bennennigreedy}.

\begin{example}
Let $n=3, d=2$ and $w=2$ then codewords from greedy search $AGG, ACC, GGC, GCA,\\ CCG, CTC$.

\end{example}

\item \textit{Lexicographic Approach:} These types of algorithms take into consideration a particular arrangement of the DNA codewords. It may be an alphabetical order or ascending order or any such kind of ordered arrangement of codes such that the arrangement of the DNA codewords satisfying the DNA constraints are ordered in a specific manner \cite{king2003bounds}.
\end{enumerate}


\item \textbf{Simulated Annealing Approach:} Simulated Annealing is a meta heuristic algorithm derived from thermodynamic principles  \cite{montemanni2014three}. These types of algorithms work on the set of codewords in which not all the codewords in the given set satisfies the constraints specified. These algorithms attempts to change the codewords with the objective of reducing the constraint violations to the constraints and try to make them feasible. The set with a feasible solution is derived when there is no violation of constraints.

\item \textbf{Stochastic Local Search Approach:} We search for the codewords with parameters $(n,d, w = n/2)$ such that the length of the codewords is $n$ and the $GC$-content is $\lfloor{n/2}\rfloor $ and the minimum Hamming distance is atleast $d$. These types of algorithms work on random set of codewords while solving the problem. It generally considers an initial set of random $k$ codewords and then removes the one which doesn't satisfy the constraints. For more details on the the algorithm, reader can refer to \cite{tulpan2003stochastic}. 

\begin{example}
Let random codewords $k=16$ then set is $AA, AT, AG, AC, GA, GT, GC, GG, CT, CA, CC,\\ CG, TA, TC, TG, TT$. Codewords with $n=2, d=2$ and $w=1$ are $GC, AG, CA, TC$. 
\end{example}

 \item \textbf{Genetic Algorithm: }In this work, genetic algorithm is used to search efficient and reliable codes by minimizing the mis-hybridization error. This codewords generated were unique in terms of Hamming distance that satisfy the Hamming bound \cite{deaton1996genetic}.
 
\item\textbf{Template Based Method:} This method was introduced in \cite{arita2002dna,liu2003dna,kobayashi2003template} that involves two step process. Initially a template is designed which is mapped to binary error correcting codes and combination of template and codeword results into DNA codewords.  This method is not optimal because the code size is limited depending of the size of error correcting code used and selection of template. Mapping of the template to the codewords is subjected to specific application.
 
\begin{example}
 Let template $t = 1001101$ and a codeword $c = 0010111$. Suppose the map define $11 \rightarrow A$, $10 \rightarrow T$, $01\rightarrow G$ and $00 \rightarrow C$ for position of $1$ and $0$ in t is $1 = [AT]$ and $0 = [GC]$ followed by position in codewords then the DNA codes will be $TCGTAGA$.
 \end{example}

\item \textbf{Algebraic Coding Approach:} By using the algebraic coding, DNA codes are constructed from fields and rings by  mapping  the elements of the field and rings to the DNA nucleotides \cite{selvakumar2014unconventional}. 

\begin{enumerate}
\item\textbf{Codes over Fields}:
\begin{enumerate}
\item \textit{Linear codes over $GF(4)$ :} 
In this approach, DNA codewords are constructed from $GF(4)$ by using different one-to-one mapping the elements of $GF(4)$ to DNA nucleotides \cite{gaborit2005linear}. The mapping is preferred from \{0,1, $\omega, \omega^2$\} to \{A,C,G,T\} with respect to the codes used. There linear code \cite{smith2011linear} and additive codes.The method used have improved the lower bounds on $GC$ constraints and extended the result on the length of DNA code to $n \leq 30$. Researchers extended this construction for non linear codes and cyclic codes \cite{aboluion2012linear}. Also DNA codes over $GF(4)$ were constructed using BCH codes in \cite{faria2010dna} in which the protein and targeting sequences are identified as codewords of error-correcting BCH codes.

\item \textit{Linear and Additive codes over $GF(4)$:} 
In this paper, DNA codes are constructed considering linear codes and additive codes over $GF(4)$ pf odd length following Hamming distance constraint and reverse complement constraint. In \cite{abualrub2006construction}, DNA codes of length $7,9,11$ and $13$ have been considered. DNA nucleotides $\{A,C,G,T\}$ have been mapped to $0,\omega,\overline{\omega}$ and $1$ respectively with $\overline{\omega}$ = $\omega^2$ and $\omega^2 + \omega + 1 = 0$. Each codeword has been mapped to a polynomial and the Trace map $ Tr : GF(4) \rightarrow GF(2)$ is stated as :
\begin{center} $Tr(x) = x + x^2$ \end{center}

\item \textit{Extended, Additive, Additive Extended Cyclic codes over $GF(4)$:}
In referred work, DNA codes satisfying $GC$-content constraint and a minimum Hamming distance constraint were constructed using computer algebra systems Magma \cite{cannon2005magma} and Maple \cite{heck1993introduction}. Longer codes of higher length $4 \leq n\leq 30$ were derived from $GF(4)$, additive codes over $GF(4)$ and $\Z_4$ (see Figure \ref{computeralgbra}). Moreover it was claimed that by using different mapping from fields or rings to DNA codewords can result into different lower bounds. Further the bounds on the DNA codes satisfying set of constraints were ameliorated by shortening and puncturing of obtained codes \cite{niema2011construction}.

\begin{figure}[ht]
\begin{center}
		\tikzstyle{block 1} = [rectangle, draw, fill=white!20, text width=14.5cm,  minimum height=1em, node distance=3cm]
		\tikzstyle{block 2} = [rectangle, draw, fill=white!20, text width=14.5cm,  minimum height=1.5em, node distance=0.8cm]
\begin{tikzpicture}
			\node [block 1] (1) {Cyclic DNA codes over $GF(4)$ have been computed for $4\leq n\leq 30$.};
			\node [block 2, below of=1] (2) {Extended cyclic DNA codes over $GF(4)$ have been computed for $4\leq n\leq 30$.};
			\node [block 2, below of=2] (3) {Additive cyclic DNA codes over $GF(4)$ have been computed for $4\leq n\leq 19$.};			
			\node [block 2, below of=3] (4) {Additive extended cyclic DNA codes over $GF(4)$ have been computed for $4\leq n\leq 20$.};			
			\node [block 2, below of=4] (5) {Cyclic DNA codes over $\Z_4$ have been computed for $4\leq n\leq 24$.};			
			\node [block 2, below of=5] (6) {Extended cyclic DNA codes over $\Z_4$ have been computed for $4\leq n\leq 24$.};			
			\node [block 2, below of=6] (7) {Cosets of cyclic DNA codes over $GF(4)$ have been computed for $4\leq n\leq 20$.};			
			\node [block 2, below of=7] (8) {Cosets of extended cyclic DNA codes over $GF(4)$ have been computed for $4\leq n\leq 20$.};			
			\node [block 2, below of=8] (9)
			{Cosets of additive cyclic DNA codes over $GF(4)$ have been computed for $4\leq n\leq 14$.};
			\node [block 2, below of=9] (10) {Cosets of additive extended cyclic DNA codes over $GF(4)$ have been computed for $4\leq n\leq 15$.};			
			\node [block 2, below of=10] (11) {Cosets of cyclic DNA codes over $\Z_4$ have been computed for $4\leq n\leq 20$.};			
			\node [block 2, below of=11] (13) {Cosets of extended cyclic DNA codes over $\Z_4$ have been computed for $4\leq n\leq 30$.};
\end{tikzpicture}
\end{center}
 \caption{DNA codes construction methods using Computer Algebra Systems such as Maple and Magma \cite{niema2011construction} .}
    \label{computeralgbra}
\end{figure}
\end{enumerate}

\item \textbf{Codes over Rings :} Algebraic construction of DNA codes was further extended to codes over rings. Different rings are used to construct DNA codes by mapping rings elements to DNA nucleotides.

\begin{enumerate}

\item \textit{DNA sequences generated by $\Z_4$ linear codes:} In \cite{rocha2010dna}, a biological coding system which modeled the existence of error correcting codes in the DNA structure. Model consists of an encoder and modulator. The encoder consist of a mapper that converts the DNA nucleotides to elements of and BCH codes over $\Z_4$) and a modulator consist of a genetic code, tRNA (transfer RNA that serves as the connecting link between the mRNA (messenger RNA) to the amino acid sequence of proteins) and ribosome (protein synthesizer of the cell) which is associated with signals that convert the genetic codons to protein. The DNA and protein coding sequences from different species have been identified as the codewords over linear codes over $\Z_4$. A class of error correcting-code BCH codes with parameters $(n,k,3)$ have been used in the encoder to construct the DNA codes over  $\Z_4$. \\

In \cite{feng2015constructions} \cite{varbanov2014method}, the self dual codes over $\Z_4$ are used for construction for DNA codes. Additionally, GC weight enumerator of the DNA codes that determines the number of Gs and Cs in the codeword. GC wright enumerator helps in the construction of DNA codes satisfying GC- content constraint. Self dual DNA codes over $\Z_4$ are developed by using mapping as $A \rightarrow 0 $ , $C \rightarrow 1$, $T \rightarrow 2$ and $G \rightarrow 3$. The following generator matrix $G$ over $\Z_4$ is considered and let $K_4$ denote the codeword over $\Z_4$ formed from $G$. There are $16$ codewords and $w_a(c)$ where $a \in \Z_4$ and $k \in K_4$ as shown in Table \ref{Table:z4}.

\begin{center}
$G=\begin{bmatrix}
    1 & 1 & 1 & 1 \\
    0 & 2 & 0 & 2 \\
    0 & 0 & 2 & 2 \\
\end{bmatrix}$
\end{center}

\begin{table}[ht]
\centering
\begin{tabular}{|l l l l l| l l l l l|}
\hline
 $K_4$ & $w_0$ & $w_1$ & $w_2$ & $w_3$ & $K_4$ & $w_0$ & $w_1$ & $w_2$ & $w_3$ \\
 \hline
 (0000)&4&0&0&0&(1111)&0&4&0&0\\
 (2222)&0&0&4&0&(3333)&0&0&0&4\\
 (0202)&2&0&2&0&(1313)&0&2&0&2\\
 (2020)&2&0&2&0&(3131)&0&2&0&2\\
 (0022)&2&0&2&0&(1133)&0&2&0&2\\
 (2200)&2&0&2&0&(3311)&0&2&0&2\\
 (0220)&2&0&2&0&(1331)&0&2&0&2\\
 (2002)&2&0&2&0&(3113)&0&2&0&2\\
\hline
\end{tabular}
\caption{$(4,16,3)$ code is generated by $G$ where $w_a(c)$ where $a \in \Z_4$ and $k \in K_4$ \cite{feng2015constructions}.}
\label{Table:z4}
\end{table}

\item\textit{Lifted Polynomials over $\F_{16}$}: In \cite{oztas2013lifted}, reversible codes by using special family of polynomials denoted as lifted polynomials over $\F_4$ which generates the reversible codes of odd length over $\F_16$ are constructed. $4$-lifted polynomial is used to generate the DNA code of even length by using the correspondence between pair of DNA nucleotides to elements of the ring $\F_16$. Table \ref{table:f16} preserves the property that if DNA pair is mapped to an element of $\F_{16}$ then reverse of that DNA pair is mapped to fourth power of the element of $\F_{16}$. For example, $\alpha^2 \rightarrow $ GC then $(\alpha^2)^4 \rightarrow $ CG.

\begin{table}[ht]
    \centering
\begin{tabular}{|L{1cm} L{2cm} L{3cm} L{2.5cm}|}
\hline
Sr.No & DNA Pair a & Multiplicative($\F_{16}$) & Additive \\
\hline
1. & AA & 0 & - \\
2. & TT & $\alpha^0$& $1$ \\
3. & AT & $\alpha^1$& $\alpha$ \\
4. & GC & $\alpha^2$& $\alpha^2$ \\
5. & AG & $\alpha^3$& $\alpha^3$ \\
6. & TA & $\alpha^4$& $1 + \alpha$ \\
7. & CC & $\alpha^5$& $\alpha + \alpha^2$ \\
8. & AC & $\alpha^6$ & $\alpha^2 + \alpha^3$ \\
9. & GT & $\alpha^7$ & $1 + \alpha + \alpha^3$ \\
10. & CG & $\alpha^8$ & $1 + \alpha^2$\\
11. & CA & $\alpha^9$ & $\alpha + \alpha^3$ \\
12. & GG & $\alpha^{10}$ & $1 + \alpha + \alpha^2$ \\
13. & CT & $\alpha^{11}$ & $\alpha + \alpha^2 + \alpha^3$ \\
14. & GA & $\alpha^{12}$ & $1 +\alpha + \alpha^2 + \alpha^3$  \\
15. & TG & $\alpha^{13}$ & $1+\alpha^2 + \alpha^3$  \\
16. & TC & $\alpha^{14}$ & $1+\alpha^3$  \\
\hline
\end{tabular}
\caption{Mapping from DNA nucleotide Pair to element of $\F_{16}$.\cite{oztas2013lifted}}
\label{table:f16}
\end{table}

\item \textit{DNA codes over $\F_2[u]/u^4-1$:} In\cite{yildiz2012cyclic}, cyclic DNA codes of odd length are obtained from $\F_2[u]/(u^4-1)=\{a+bu+cu^2+du^3\mid a, b, c, d \in \F_2\}$ where $u^4 = 1$ commutative ring is considered. All its ideals are listed here $\langle 0 \rangle = \langle(1+u)^4 \rangle \subset \langle (1+u)^3 \rangle  \subset \langle (1+u)^2 \rangle \subset \langle (1+u) \rangle \subset \R$. The $16$ elements of ring $\R$ are mapped to $2$-length nucleotides. The following mapping shown in Table \ref{table:u4} was considered in the paper\cite{yildiz2012cyclic}. 
\begin{table}[ht]
\centering
 \begin{tabular}{|l l| l l |l l |l l |} 
 \hline
 Element & Map & Element & Map & Element & Map & Element & Map\\
 \hline
 AA & 0 & AT & 1 + u & GT & 1 & CT & 1+ u + $u^2$ \\ 
 TT & 1 + u + $u^2$ + $u^3$ & TA & $u^2$ + $u^3$ & TG & $u^2$ & TC & 1 + $u^2 + u^3$ \\ 
 GG & 1 + $u^2$ & GC & $u + u^2$ & AC & $1+ u + u^3$ & AG & u \\ 
 CC & $u + u^3$ & CG & $1 + u^3$ & CA & $u + u^2 + u^3$ & GA & $u^3$ \\ 
 \hline
\end{tabular}
\caption{Mapping from DNA Nucleotide Pair to elements of the Ring $\F_{2}+ u\F_2 + u^2\F_2 + u^3\F_2$ where $u^4=1$ \cite{yildiz2012cyclic}.}
\label{table:u4}
\end{table}
This mapping preserved the complementary and reverse property by adding $1+u+u^2+u^3$ and multiplying $u^2$ respectively. 
To find complement of $AA$, we add $1+u+u^2+u^3$ to $0$, it will give $1 + u + u^2 + u^3$ = $TT$. To find reverse $AA$, multiply $0$ to $u^2$ will result in $0$ =$AA$. 

 In \cite{guenda2013cyclic} DNA cyclic codes of arbitrary length satisfying the reverse complement constraint are constructed by using additive stem distance. The correspondence between the ring elements and DNA is established by following mapping mentioned in Table \ref{u4}. this preserves the reverse complement property of the DNA codewords by the $x + x^c = u^3 + u^2 + u+1$. The reverse of the DNA code is obtained by multiplying $u^2$ to any element $x$ of the ring $\R$.

\begin{table}[ht]
\centering
 \begin{tabular}{|l l| l l |l l |l l |} 
 \hline
 Element & Map & Element & Map & Element & Map & Element & Map\\
 \hline
 GG & 0 & AT & 1 + u & GT & 1 & CT & 1+ u + $u^2$ \\ 
 CC & 1 + u + $u^2$ + $u^3$ & TA & $u^2$ + $u^3$ & TG & $u^2$ & TC & 1 + $u^2 + u^3$ \\ 
 GC & 1 + $u^2$ & AA & $u + u^2$ & AC & $1+ u + u^3$ & AG & u \\ 
 CG & $u + u^3$ & TT & $1 + u^3$ & CA & $u + u^2 + u^3$ & GA & $u^3$ \\ 
 \hline
\end{tabular}
\caption{Mapping from DNA Nucleotide Pair to elements of the Ring $\F_{2}+ u\F_2 + u^2\F_2 + u^3\F_2$ where $u^4=1$ \cite{yildiz2012cyclic}.}
\label{u4}
\end{table}

\item \textit{DNA cyclic codes over $\F_2 + u\F_2$  where $u^{2}=0$:} DNA codes of even length following reverse and reverse complement constraints have been studied in \cite{liangcyclic}. The field $\F_2$ is a subring of $\R$. A linear code $C$ of length $n$ over $\R$ is defined to be an additive submodule of the $\R$-module $\R^n$. A cyclic code of length $n$ over $\R$ is a linear code with the property that if $\left( c_{0}, c_{1}, \ldots , c_{n−1} \right) \in C$ then $\left( c_{n−1}, c_{0}, \ldots , c_{n−2} \right)\in C$. 
An $n$-tuple $c = (c_0, c_1, \ldots , c_{n−1}) \in \R^{n}$ is identified with the polynomial $c_{0} + c_{1}x + \ldots + c_{n−1}x^{n−1}$ in the ring $\R_{n} = \R[x]/\left( x^{n} − 1 \right)$, which is called the polynomial representation of $c = \left( c_{0}, c_{1}, . . . , c_{n−1} \right)$. Here $\R$ = $\F_2 + u\F_2$ with $\{0,u,1,1+u\}$ elements are in one to one correspondence with nucleotides $ A,T,G$ and $C$ such that $0 \rightarrow A$, $u \rightarrow T$, $u + 1 \rightarrow C$ and $1 \rightarrow G$. The DNA codes of length $8$ and $10$ are obtained from $ C = g^{6} + u \left( x^{5} + x \right)$ and $ C = g_{1} g_2^2 + ug_2, ug_1 g_2$. Necessary and sufficient conditions for cyclic codes to follow the reverse and reverse-complement properties have also have been studied. To preserve the reverse complement constraint o find complement of A, we add u to $0$, it will give $u$ = T.

 \item \textit{DNA codes over $\Z_4 + u\Z_4$:} DNA cyclic codes of odd lengths following reverse and reverse complement constraint are constructed in \cite{pattanayak2015cyclic}. Here ring $\Z_4 + u\Z_4 = {a + ub \mid a, b \in \Z_4}$ with $u^2 = 0$ is considered. Reversible and cyclic reversible complement codes are discovered in this paper. In this work defined a Gray map that allows them to translate the properties of DNA codes to binary codewords $i.e.$ $\phi : \Z_4 + u\Z_4 \rightarrow \Z_4^{2}$ such that $\phi (a+ub) = (b,a+b)$ where $a,b \hspace{0.2cm} \in \hspace{0.2cm} \Z_4$. In this paper, $16$ pairs of nucleotides which are mapped as shown in Table \ref{table:z4}.
 
\begin{table}[ht]
\centering
 \begin{tabular}{|l l| l l| l l| l l|} 
 \hline
 Element & Map & Element & Map & Element & Map & Element & Map\\
 \hline
 AA & 0 & TT & 1+u & GG & 1 & CC & u \\ 
 AT & 2 & TA & 3+u & GC & 3 & CG & 2+u \\ 
 GT & 2u & CA & 1+3u & AC & 3u & TG & 1+2u \\ 
 CT & 2+3u & GA & 3+2u & AG & 2+2u & TC & 3+3u \\  
\hline
\end{tabular}
\caption{Mapping from DNA nucleotide pair to element of the Ring $\Z_4 + u\Z_4$ where $u^2=0$\cite{pattanayak2015cyclic}.}
\label{table:z4}
\end{table}

\item \textit{$\F_4 [u]/<u^2+1>$ where $u^2 =1$:} In \cite{ma2015cyclic} self-reciprocal complement cyclic codes from $\R$ with $\F_4 = \{0,1,\alpha,\alpha + 1\}$ where $\alpha$ is the root of primitive polynomial $x^2+x+1$ over $\F_2$ are studied. The DNA code of specific length $6$ over the ring is considered. One to one correspondence is establish between pairs of nucleotides and $16$ elements of the ring. DNA cyclic codes constructed followed reverse complement, $GC$ content and Hamming distance constraints.  The basis is $\{1,u+1\}$ and then every element of $\R$ is expressed in the form of $a+b(u+1)$ where a,b $\in$ $\F_4$. The mapping is done as in Table  \ref{table:f4_u2}

\begin{table}[ht]
\centering
\begin{tabular}{|l l| l l| l l| l l|} 
 \hline
 Element & Map & Element & Map & Element & Map & Element & Map\\
 \hline
 AA & $0$ & TT & $\alpha +1 (\alpha +1)u$ & GT & $1$ & CA & $\alpha+(\alpha+1)u$ \\ 
 AG & $\alpha$ & TC & $1+(\alpha+1)u$ & AT & $\alpha+1$ & TA & $(\alpha+1)u$ \\ 
 TG & $u$ & AC & $\alpha+1+\alpha u$ & GA & $\alpha u$ & CT & $1+\alpha+u$ \\ 
 GC & $1+\alpha u$ & CG & $\alpha+u$ & CC & $1+u$ & GG & $\alpha+\alpha u$ \\  
\hline
\end{tabular}
\caption{Mapping between the pair of DNA nucleotides and Ring $\F_4$ \cite{ma2015cyclic}.}
\label{table:f4_u2}
\end{table}

\item\textit{DNA codes from $\F_2 + u\F_2 + v\F_2 + uv\F_2 $ with $u^2 = 0$ and $v^2=v$:} The structure of cyclic DNA codes of an arbitrary length over $\R_2 = \F_2 + u\F_2 + v\F_2 + uv\F_2$ was studied and the relation to codes over $\R_1=\F_2+u\F_2$ by defining Gray map between $\R_2$ and $\R_1^2$ was established \cite{zhu2015cyclic}. DNA codes following reverse, reverse complement constraints are studied. The Gray map from $\R_2$ to $\R_1$ is defined as $\phi(a + bv) = (a, a + b)$. $GC$ weight over the ring was also introduced  by using image of Gray map. One type of nontrivial automorphisms can be defined over $\R_2$ as follows : $\sigma : \F_2 + u\F_2 + v\F_2 + uv\F_2 \rightarrow  \F_2 + u\F_2 + v\F_2 + uv\F_2,$ \\
$a + bv \rightarrow a + (1 + v)b$ such that $a, b \in F_2 + uF_2$. Table is defined using : $\Phi\left( c \right)  : C \rightarrow S^{2n}_{D4}$, \\
$\left( a_0+b_{0}v, a_1+b_{1}v, \dots , a_{n−1}+b_{n−1}v \right)  \mapsto \left( a_0, a_1, \dots , a_{n−1}, a_0+b_0, a_1+b_1, \dots , a_{n−1}+b_{n−1} \right) $.
Below is the Table \ref{table:f2uf2uvf2} for mapping elements of the ring to DNA described in the paper. For instance, $(c0, c1, c2, c3)$ = $(u+v, u, v, 1)$ is mapped to $T C T T A G T C G G $.

\begin{table}[ht]
\centering
\begin{tabular}{|L{2cm}| L{1.5cm}| L{1.5cm}|L{2cm}| L{1.5cm}| L{2cm}|}
\hline
Elements a & Gray Images & Double DNA Pairs $\zeta(a)$ & Elements a & Gray Images & Double DNA Pairs $\zeta(a)$  \\
\hline
 0 & (0,0) & AA & v & (0,1) & AG \\
 uv & (0,u) & AT &  v+uv & (0,1+u) & AC \\
 1 & (1,1) & GG &  1+v & (1,0) & GA \\
 1+uv & (1,1+u) & GC & 1+v+uv & (1,u) & GT \\
 u & (u,u) & TT &  u+v & (u,1+u) & TC \\
 u+uv & (u,0) & TA &  u+v+uv & (u,1) & TG \\
 1+u & (1+u,1+u) & CC & 1+u+v & (1+u,u) & CT \\
 1+u+uv & (1+u,1) & CG & 1+u+v+uv & (1+u,0) & CA \\
\hline
\end{tabular}
\caption{Mapping between DNA pair and elements of the Ring $\F_2 + u\F_2 + v\F_2 + uv\F_2 $ with $u^2 = 0$ and $v^2=v$ \cite{zhu2015cyclic}.}
\label{table:f2uf2uvf2}
\end{table}

\item \textit{codes over $\F_4+v\F_4$:} In linear, constacyclic and cyclic codes over the ring $\R = F_4[v]/(v^2-v)$ are constructed in \cite{bayram2015codes}. The ring $\R= \{a + vb \vert a,b \in \F_4\}$ is non-chain finite semi-local Frobenius ring with $16$ elements. The $4$ elements are $\F_4 = \{0,1,\omega , \omega + 1\}$ where 
$\omega^2 = \omega + 1.$ The Gray map from $\R$ to $\F_4 \times \F_4$ is given by :
\begin{center} 
$\phi (c) = (a + b, a)$. 
\end{center}
If $a = (a_1, a_2,\ldots , a_n) \in \R_n$, then the Hamming weight of $a$ is the sum of the Hamming weights of its components, $i.e. w(a) = \sum_{i=1}^{n}w(a_i)$. The Hamming distance between a and b in $\R$ is $d(a, b) = w(a − b)$. The Lee weight of any element of R is the Gray image of its Hamming weight, $i.e.$ $w_L(c) = w_H(\phi(c))$. Below is the Table \ref{table:f4vf4} for mapping used in \cite{srinivasulu2015reversible}.
\begin{table}[ht]
\centering
\begin{tabular}{|L{1.5cm}  L{1.5cm}  L{2cm}| L{2cm}  L{2.2cm}  L{2cm}|}
\hline
 Elements & Gray Images & Double DNA Pairs $\xi(a)$ &  Elements & Gray Images & Double DNA Pairs $\xi(a)$ \\
\hline
 $0$ & $(0,0)$ & AA & $1$ & $(1,1)$ & TT \\
  $\omega$ & $(\omega,\omega)$ & CC &  $1 + \omega$ & $(1+\omega,1+\omega)$ & GG \\
 $v$ & $(1,0)$ & TA &  $1+v$ & $(0,1)$ & AT \\
$v + \omega$ & $(1+\omega,\omega)$ & GC & $1+v+\omega$ & $(\omega,1+\omega)$ & CG \\
 $v\omega$ & $(\omega,0)$ & CA &  $1+v\omega$ & $(1+\omega,1)$ & GT \\
 $\omega+v\omega$ & $(0,\omega)$ & AC &  $1+\omega+v\omega$ & $(1,1+\omega)$ & TG \\
 $v+v\omega$ & $(1+\omega,0)$ & GA &  $1+v+v\omega$ & $(\omega,1)$ & CT \\
 $\omega+v+v\omega$ & $(1,\omega)$ & TC &  $1+v+\omega +v\omega$ & $(0,1+\omega)$ & AG \\
\hline
\end{tabular} 
\caption{Mapping between DNA Pair and elements of the Ring $\F_4 + v\F_4$\cite{bayram2015codes}.}
\label{table:f4vf4}
\end{table}

\item \textit{$\R = \F_2[u]/(u^6)$:} In  \cite{bennenni2015new}, DNA cyclic codes over a family $\R_1 = \F_2[u] / (u^6)$ and ring $\R_2 = \F_2 + u\F_2$ where $v^2=v$ satisfying reverse complement constraint have been constructed. In this a new family of DNA skew cyclic codes is introduced over ring $\R = \F_2 + v\F_2 = {0, 1, v, v + 1}$ where $v^2 = v$.
The ring $\R_1 = \F_2[u] / (u^6) = \{a_0 + a_1u + a_2 u^2 + a_3 u^3 + a_4 u^4 + a_5 u^5; a_i \in \F_2 , u^6 = 0\}$.
There is direct map between $64$ elements of the ring to $64$ codons (three nucleotides) used in nature as a substrate for aminoacid synthesis shown in Table \ref{table:f2u6}.

\begin{table}[ht]
\centering
\small
\begin{tabular}{|L{0.7cm} L{3cm} |L{0.7cm} L{2cm}| L{0.7cm} L{2.7cm}| L{0.7cm} L{2.7cm} |}
\hline
DNA Codons & Ring Element & DNA Codons & Ring Element & DNA Codons & Ring Element & DNA Codons & Ring Element \\
\hline
 & & & & & & & \\
CCC & $u^5+u^4+u^3+u^2+u+1$ & GGG & $0$ & ACT & $u^3+u^2+u+1$ & GTC & $u^4+u^2+u+1$ \\
GGA & $u^5+u^4+u^3+u^2+u$   & CCT & $1$ & ACG & $u^3+u^2+u$   & ACA & $u^3+u^2+u+1$ \\
GGC & $u^5+u^4+u^3+u^2+1$   & CCG & $u$ & TTT & $u^4+u^2+1$   & GAC & $u^5+u^3+u^2+1$ \\
GGT & $u^5+u^4+u^3+u^2$   & CCA & $u+1$ & TTG & $u^4+u^2+u$   & AGG & $u^5 + u^3+u+1$ \\
AGG & $u^5+u^4+u^3+u+1$ & TCC & $u^2$ & CTA & $u^4+u+1$ & GAT & $u^5+u^3+u^2$ \\
CGG & $u^5+u^4+u^2+u+1$ & GCC & $u^3$ & GTT & $u^4+u^3+1$ & GTA & $u^4+u^3+u+1$ \\
GAG & $u^5+u^3+u^2+u+1$ & CTC & $u^4$ & GTG & $u^4+u^3+u$ & ATT & $u^4+u^3+u^2+1$ \\
AGA & $u^5+u^4+u^3+u^2+u$ & TCT & $u^2+1$ & TCA & $u^2+u+1$ & ATA & $u^4+u^3+u^2+u$ \\
AGC & $u^5+u^4+u^3+1$ & TCG & $u^2+u$ & CAA & $u^5+u^2+u$ & ATC & $u^4+u^3+u^2$ \\
ATG & $u^4+u^3+u^2+u+1$ & TAC & $u^5$ & CAC & $u^5+u^2+1$ & TGA & $u^5+u^4+u$ \\
AGT & $u^5+u^4+u^3$ & TAT & $u^5+1$ & GCA & $u^3+u+1$ & AAT & $u^5+u^2+u+1$ \\
CGA & $u^5+u^4+u^2+u$ & GCT & $u^3+1$ & TTA & $u^4+u^3$ & AAA & $u^5+u^3+u$ \\
CGC & $u^5+u^4+u^2+1$ & GCG & $u^3+u$ & ACC & $u^3+u^2$ & TGC & $u^5+u^4+1$ \\
CGT & $u^5+u^4+u^2$ & TAA & $u^5+u$ & CAT & $u^5+u^2$ & AAC & $u^5+u^3+1$ \\
TGG & $u^5+u^4+u+1$ & CTG & $u^4+u$ & TGT & $u^5+u^4$ & TCC & $u^4+u^2$ \\
GAA & $u^5+u^4+u^3+u^2+u$ & CTT & $u^4+1$ & CAG & $u^5+u^3$ & TAG & $u^5+u+1$ \\
\hline
\end{tabular}
\caption{Mapping between $64$ elements of the Ring $\F_2 +u\F_2 +u^2\F_2 + u^3\F_2 + u^4\F_2 + u^5\F_2$ and $64$ codons\cite{bennenni2015new}.}
\label{table:f2u6}
\end{table}

\item \textit{DNA codes over ring $\F_2[u]/u^2-1$ :} Ring $\R = \F_2[u]/u^2-1 = \{0,1,u,1+u\}$ where $u^2=1$ was described \cite{siap2009cyclic} \cite{mostafanasab2016cyclic} . Elements of the ring were directly mapped to DNA nucleotides such that reverse complement constraint is conserved. By adding $1+u$ to elements of rings, complement can be obtained. For example $0 \rightarrow A, u \rightarrow C, 1 \rightarrow G, 1+u \rightarrow T$ preserves the complement property: $\overline{0} = 1 + u$ and $\overline{u} = 1$. \\
The elements of the ring $\R$ can be mapped to the elements of $F_2 = \{0,1\}$ via the map $\theta$ where 
$\theta(0) = \theta(u+1) = 0$ and $\theta(1) = \theta(u)=1$. Let C be a cyclic code in $\R_n$. It can be extended to the map $\theta$ to a map $\varphi : C \rightarrow Z_2[x]/(x^n-1)$ defined by
$\varphi(a_0 + a_{1}x + a_{2}x^{2} + \ldots + a_{n-1}x^{n-1}) = \theta(a_0) + \theta(a_1)x + \ldots +\theta(a_{n-1}x^{n-1})$.

\item \textit{DNA cyclic codes over $\F_2 + u\F_2$} : In \cite{guenda2012construction} odd length codes over rings satisfying reverse complement, $GC$-Content and thermodynamic constraints are studied. They are obtained from the cyclic complement reversible code. Infinite family of BCH DNA codes are constructed. The mapping $\Phi$ called Gray Map has been used to map linear codes over $\R$ to binary linear codes. 
The Gray map $\Phi$ is the distance-preserving map $ (R^n,$ Lee distance) $\rightarrow ( F^{2n}_2$ ,Hamming distance). 

\item \textit{Cyclic DNA codes over $\Z_4 + w \Z_4$} : Recently, cyclic DNA codes over $R = \Z_4 + w\Z_4$ where $w^2 = 2$ and $S = \Z_4 + w\Z_4 + v\Z_4 + wv\Z_4$ where $v^2 = vw$ have been discovered \cite{dertli2016cyclic}. In this work, odd length DNA cyclic codes over $R$ satisfying reverse and reverse complement constraint is studied . Also, a family of DNA skew cyclic codes with reverse complement property over R is constructed. Binary images of the cyclic DNA codes over $R$ and $S$ is determined. The correspondence between elements of ring $R$ and double DNA pairs are establish as described in the Table \ref{table:z4wz4}.

\begin{table}[ht]
\centering
\begin{tabular}{|L{1.5cm}  L{1.5cm}  L{2cm}| L{2cm}  L{2.2cm}  L{2cm}|}
\hline
 Elements & Gray Images & Double DNA Pairs $\xi(a)$ &  Elements & Gray Images & Double DNA Pairs $\xi(a)$ \\
\hline
 $0$ & $(0,0)$ & AA & $1$ & $(1,0)$ & CA \\
  $2$ & $(2,0)$ & GA &  $3$ & $(3,0)$ & TA \\
 $\omega$ & $(0,1)$ & AC &  $2\omega$ & $(0,2)$ & AG \\
$3\omega$ & $(0,3)$ & AT & $1+\omega$ & $(1,1)$ & CC \\
 $1+2\omega$ & $(1,2)$ & CG &  $1+3\omega$ & $(1,3)$ & CT \\
 $2+\omega$ & $(2,1)$ & GC &  $2+2\omega$ & $(2,2)$ & GG \\
 $2+3\omega$ & $(2,3)$ & GT &  $3+\omega$ & $(3,1)$ & TC \\
 $3+2\omega$ & $(3,2)$ & TG &  $3+3\omega$ & $(3,3)$ & TT \\
\hline
\end{tabular} 
\caption{Mapping between DNA Pair and elements of the Ring.}
\label{table:z4wz4}
\end{table}

\end{enumerate}
\end{enumerate}

\item\textbf{Algebraic Number Theory codes:} Aforementioned all the methods include construction of DNA codes from classical coding theory and heuristic approaches works well for small length $n$. In this author constructed the DNA codes using algebraic number theory \cite{hong2016construction}, making the first attempt, by using irreducible cyclic codes to built DNA codes with large $n$ ($<1000$) and number of codewords $(M < 7000)$ satisfying the $GC$ content constraint.
 

\end{enumerate}

\section{Software tools for DNA Codes Generation}
There are different tools developed for designing the DNA codewords namely DNA sequence Generator \cite{feldkamp2003software} and evolutionary algorithm based program-PUNCH (Princeton University Nucleotide Computing Heuristic) \cite{ruben2001punch} were used to find set of dissimilar sequences. DNA sequence generator and compiler use graph based approach based on the overlapping sub-sequences. The GUI of the software allows the user to import the the DNA sequence to the sequence wizard with different parameters. It also calculate the melting temperature of the DNA sequences. It check for the reverse complement constraints and forbidden DNA strands. However this software do not take care of secondary structure formation of the DNA strands. PUNCH is used for performing various DNA computing by bit set selection. It works on randomization by selecting three basic parameters N, B and V where N is number of bits in the problem, B is the number of nucleotides in each bit, and V is number of variation on each bit set.

Here the author has created a web application in which the user has to first select the mode which is either specific based or range. For both modes, the user has to select the specific constraints he wants DNA codes to follow but for specific the input has parameters n and d where n is the length of the codewords and d is the Hamming distance. For range, the input parameters are n1 and n2 where n1 is the starting length and n2 is the ending length and also are d1 and d2 where d1 is the starting Hamming distance and d2 is the ending Hamming distance. The web portal will display the number of codewords and also those codewords that satisfy the constraint.

\section{Bounds on DNA Codes}
There are several bounds studied for DNA codes. This section comprehend the types of bounds obtained on set of constraints. In the Table \ref{table:bounds123}, columns are ticked for which respective bounds on constraints are obtained. Note that $n$ is the length, $d$ is minimum Hamming distance and $w$ is $GC$ content of the DNA code. One can observe that almost all the methods have obtained lower bounds on reverse constraint. But most important part is to investigate that there is no bounds on DNA codes satisfying the set of HD, $GC$, R and RC constraints altogether. Also there are very few attempts made to explore the bounds on thermodynamic constraints. One can explore the methods which allows the formulation of bounds on the thermodynamic constraints. More details on the bounds are described in Appendix \ref{App:AppendixA}.

\begin{table}[ht]
\centering
\begin{tabular}{|l| l | l | l | l| l| l| l|l | l | l | l | l | l | l | l | l | l |}
\hline 
& $1$ & $2$ & $3$ & $4$& $5$ & $6$ & $7$ & $8$& $9$& $10$ & $11$ &$12$& $13$ &$14$& $15$ & $16$& $17$\\
\hline
& & & & &   & & & & &   & & & & &   & & \\
${A}_{4}^{HD}(n,d,w)$& & & & &   & & & & &   & & & & &   & & $\checkmark$\\ \hline
& & & & &   & & & & &   & & & & &   & &  \\
${A}_{4}^{R}(n,d,w)$ & $\checkmark$ &  $\checkmark$ & & &   & & &$\checkmark$ & $\checkmark$ & &$\checkmark$ &$\checkmark$ &$\checkmark$ & &
$\checkmark$ &$\checkmark$ & $\checkmark$  \\ 
\hline
& & & & &   & & & & &   & & & & &   & &  \\ 
${A}_{4}^{RC}(n,d,w)$ & & & & &   & & & & $\checkmark$ &    & $\checkmark$ & & & &    & &  \\ 
\hline
& & & & &   & & & & &   & & & & &   & &  \\ 
${A}_{4}^{GC}(n,d,w)$ & $\checkmark$ & & $\checkmark$ & &  $\checkmark$ & & & $\checkmark$ & & $\checkmark$  & & & & &   & &  \\ 
\hline
& & & & &   & & & & &   & & & & &   & &  \\ 
${A}_{4}^{R,RC}(n,d,w)$ & & & & &   & & & & &   & & & & &   & &  \\ 
\hline
& & & & &   & & & & &   & & & & &   & &  \\ 
${A}_{4}^{RC,GC}(n,d,w)$ & & $\checkmark$ & &$\checkmark$ &  & $\checkmark$& $\checkmark$ & $\checkmark$ & &  $\checkmark$ & & & & &   & &  \\ 
\hline
& & & & &   & & & & &   & & & & &   & &  \\ 
${A}_{4}^{R,GC}(n,d,w)$ & & & & $\checkmark$ &   & & & $\checkmark$ & &   & & & & &   & &  \\ 
\hline
& & & & &   & & & & &   & & & & &   & &  \\ 
${A}_{4}^{R,RC,GC}(n,d,w)$ & & & & &   & & & & &   & & & & &   & &  \\ 
\hline
\end{tabular}
\caption{Bounds on DNA codes: 1-Johnson type Bound \cite{ignatova2008dna} \cite{king2003bounds}, 2-Halving Bounds \cite{ignatova2008dna} \cite{king2003bounds}, 3-Gilbert-Type Bounds \cite{king2003bounds}, 4- Relation between ${A}_{4}^{GC,RC}$ and ${A}_{4}^{GC,R}$ \cite{king2003bounds}, 5- ${A}_{4}^{GC}$ bound, 6- ${A}_{4}^{GC,RC}$ bound, 7- ${A}_{4}^{GC}(n,d,w)$ bound, 8- Product Bounds \cite{ignatova2008dna} \cite{king2003bounds}, 9- Relation between ${A}_{4}^{RC}$ and ${A}_{4}^{R}$ \cite{ignatova2008dna}, 10 to 16 - Relation between ${A}_{4}^{GC}$, ${A}_{4}^{RC}$ and ${A}_{4}^{R}$ \cite{marathe2001combinatorial}, 17-V. Phan bound \cite{zhang2011bounds} satisfying Hamming distance constraint. }
\label{table:bounds123}
\end{table}

\section{Applications of DNA Codes}
\begin{center}
\centering
\includegraphics[scale=0.3]{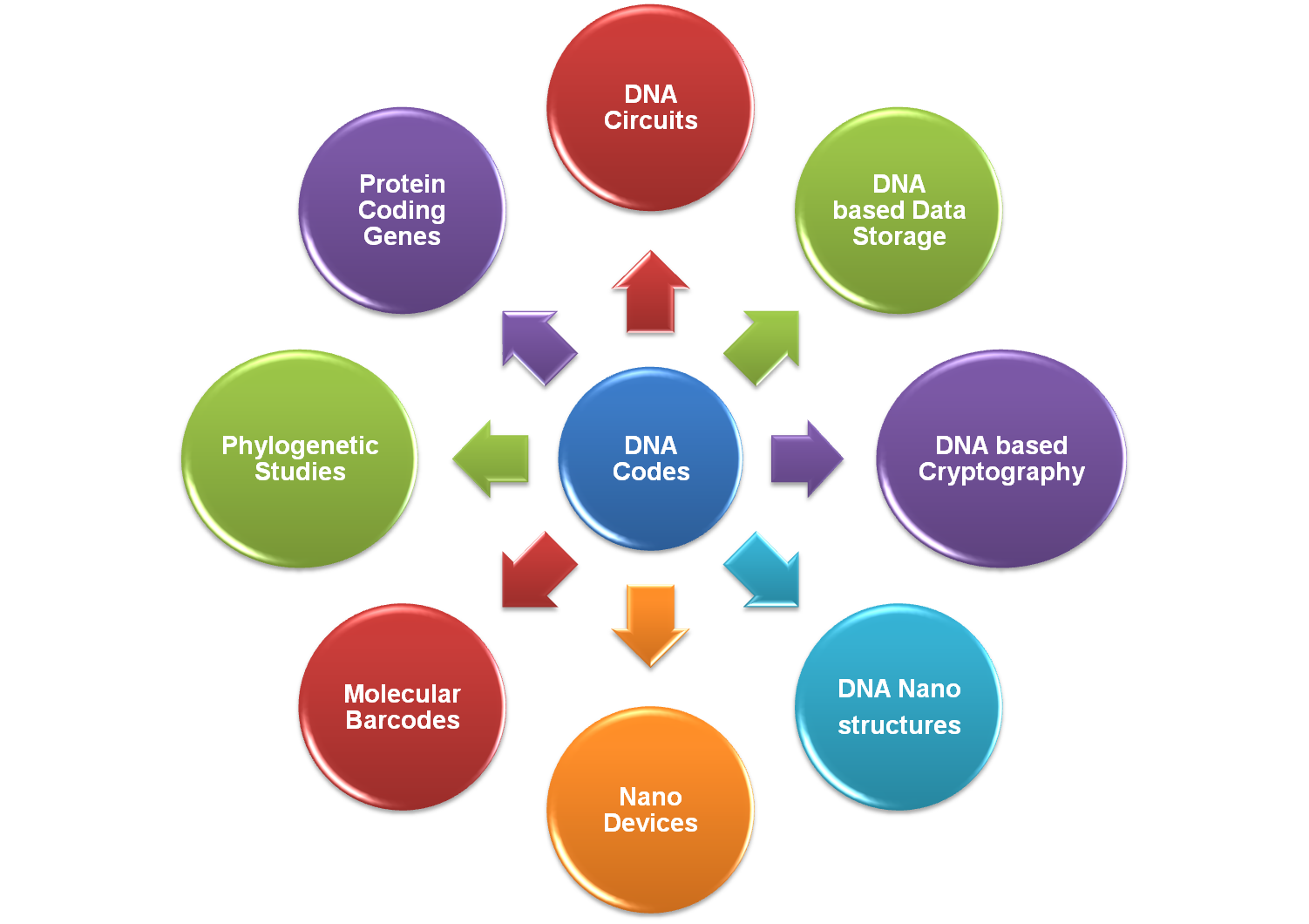}
\end{center}
DNA codes are used in various technologies like DNA computing \cite{adleman1994molecular}, surface based DNA computation \cite{smith1998surface,frutos1997demonstration,liu2000dna,wu2001improved}, DNA Microarray technology \cite{schena1995quantitative}, Molecular barcodes for chemical libraries \cite{brenner1992encoded}, DNA nanostructures \cite{reif2012self,yang2015dna}, DNA origami \cite{rothemund2006folding}, data encrytption \cite{jacob2013dna,tulpan2013hyden} \cite{aich2015symmetric,wei2012novel,enayatifar2014chaos}, data storage \cite{arita2004method,church2012next,goldman2013towards,limbachiya2015natural,yazdi2015dna}, signal processing \cite{tsaftaris2004dna} and DNA nanodevices and circuits \cite{qian2011scaling}. Recently, it has reported potential of DNA codes in phylogenetic studies \cite{garzon2011dna}. Also it has contribution in understanding gene regulatory networks \cite{dingel2008coding}, protein coding genes \cite{arques1996complementary,arques1997code} and studying the structure of genes via circular codes \cite{michel20082006}.

In DNA computing, DNA codes with specific properties are required to perform various parallel and logical operations \cite{garzon2012theory}. Molecular barcodes \cite{bystrykh2012generalized} generated from DNA codes are used as biomarkers for authentication of the products. DNA codes are used in creating DNA nanostructures that are used in potential applications like targeted drug delivery systems. DNA codes with specific properties with high stability and robustness are required for nano structures which can be achieved by using efficient encoding procedures for DNA codes. Recently, DNA is used in data hiding techniques for encryption of data more effectively. DNA codes used in this are designed as encryption or decryption keys. In last few years DNA based data storage systems have \cite{yazdi2015rewritable} received attention by many researchers. DNA codes used for data storage must have feasible property that achieve dense data storage capacity and better error correction capacity. 

To use DNA codes for any application, fundamental constraints mentioned for DNA codes are unavoidable. DNA code design must follow the constraint for stability and robustness but to design the DNA codes for specific application, required constraints must be added to DNA codes to make it more functional and practical. For instance, correlated and uncorrelated constraint was added to DNA codes for development random and re-writable DNA based data storage system. Looking at the potential of DNA codes and advancement in the biotechnology methods, DNA codes promises application in emerging technologies. 
\section{DNA Codes Table}
Many tables  on lower bounds of DNA codes satisfying set of constraints are obtained. In Table \ref{boundtable1} and \ref{boundtable2} lower bounds on DNA codes satisfying $GC$ and reverse complement (RC) constraints are mentioned. In Table \ref{RCbound}  lower bounds on DNA codes satisfying Hamming distance and reverse complement constraint. These bounds are compiled from \cite{niema2011construction} \cite{tulpan2014thermodynamic} \cite{montemanni2014three} \cite{gaborit2005linear} \cite{varbanov2014method} \cite{hong2016construction}.

\begin{center}
\begin{footnotesize}
\begin{sidewaystable}
\centering

\begin{tabular}{|l|l|l|l|l|l|l|l|l|}
\hline
n/d	&	2	&	3	&	4	&	5	&	6	&	7	&	8	&	9	\\ \hline
4	&	24	&	6	&	2	&		&		&		&		&		\\
5	&		&	15	&	3	&	1	&		&		&		&		\\
6	&	320	&	43	&	16	&	4	&	2	&		&		&		\\
7	&	135	&	256	&	35	&	11	&	2	&	1	&		&		\\
8	&		&	528	&	128	&	28	&	22	&	2	&	2	&		\\
9	&		&	1354	&	273	&	65	&	19	&	8	&	2	&	1	\\
10	&	64512	&	4542	&	860	&	210	&	54	&	17	&	8	&	2	\\
11	&		&	14405	&	2457	&	477	&	117	&	37	&	14	&	5	\\
12	&	946176	&	59136	&	14784	&	1848	&	924	&	87	&	29	&	12	\\
13	&		&	167263	&	27376	&	3974	&	924	&	206	&	62	&	23	\\
14	&		&	768768	&	192192	&	11878	&	3712	&	796	&	208	&	49	\\
15	&		&	1646240	&	411821	&	25670	&	6648	&	1600	&	410	&	109	\\
16	&		&	13174400	&	3293600	&	55376	&	55424	&	13856	&	6476	&	243	\\
17	&		&	26355520	&	6587200	&	97520	&	97450	&	12864	&	6060	&	579	\\
18	&		&	44933184	&	11232288	&	699624	&	738772	&	43632	&	43632	&	2691	\\
19	&		&	47102080	&	23647760	&	738772	&	738772	&	92252	&	11542	&	3678	\\
20	&		&	756760576	&	189432064	&	11822368	&	11806240	&	738520	&	368504	&	11452	\\
21	&		&	90291264	&	188416000	&	22573824	&	1412068	&	176772	&	45112	&	11148	\\
22	&		&	10602158336	&	2650495232	&	22607872	&	5643456	&	176772	&	353496	&	88424	\\
23	&		&	1384513088	&	670222080	&	43264648	&	10816624	&	2703694	&	676312	&	169182	\\
24	&		&	177279886336	&	44319794176	&	346436544	&	43355616	&	21631400	&	5406464	&	1351616	\\
25	&		&	21300369664	&	11204500480	&	10399676	&	1299844	&	41600552	&	10399676	&	2599688	\\
26	&		&	2532157069312	&	633038608384	&	326893568	&	618544192	&	38656528	&	83204800	&	20801200	\\
27	&		&		&	158680788992	&		&	20057442	&		&	40114884	&	40114884	\\
28	&		&	42061705248768	&	10515426312192	&	5154680832	&	10262347776	&	641396736	&	159987712	&	627776	\\
29	&		&		&	2625500086272	&		&	6691200	&		&	3853632	&		\\
30	&		&	609973884610560	&	152493461268480	&	80766566400	&	149011451520	&	9313176480	&	2332609440	&	9080016	\\
\hline
\end{tabular}
\label{boundtable1}
\caption{Lower bounds on DNA codes satisfying GC and Reverse Complement Constraints for $4 \leq n \leq 30$ and $2 \leq d \leq 9$} 
\end{sidewaystable}
\end{footnotesize}

\begin{footnotesize}
\begin{sidewaystable}
\begin{tabular}{|l|l|l|l|l|l|l|l|l|l|l|l|l|l|l|l|l|l|l|l|l|l|}
\hline
n/d	&	10	&	11	&	12	&	13	&	14	&	15	&	16	&	17	&	18	&	19	&	20	&	21	&	22	&	23	&	24	&	25	&	26	&	27	&	28	&	29	&	30	\\ \hline
4	&		&		&		&		&		&		&		&		&		&		&		&		&		&		&		&		&		&		&		&		&		\\
5	&		&		&		&		&		&		&		&		&		&		&		&		&		&		&		&		&		&		&		&		&		\\
6	&		&		&		&		&		&		&		&		&		&		&		&		&		&		&		&		&		&		&		&		&		\\
7	&		&		&		&		&		&		&		&		&		&		&		&		&		&		&		&		&		&		&		&		&		\\
8	&		&		&		&		&		&		&		&		&		&		&		&		&		&		&		&		&		&		&		&		&		\\
9	&		&		&		&		&		&		&		&		&		&		&		&		&		&		&		&		&		&		&		&		&		\\
10	&	2	&		&		&		&		&		&		&		&		&		&		&		&		&		&		&		&		&		&		&		&		\\
11	&	2	&	1	&		&		&		&		&		&		&		&		&		&		&		&		&		&		&		&		&		&		&		\\
12	&	4	&	2	&	2	&		&		&		&		&		&		&		&		&		&		&		&		&		&		&		&		&		&		\\
13	&	10	&	4	&	2	&		&		&		&		&		&		&		&		&		&		&		&		&		&		&		&		&		&		\\
14	&	21	&	8	&	4	&	2	&	2	&		&		&		&		&		&		&		&		&		&		&		&		&		&		&		&		\\
15	&	37	&	18	&	20	&	3	&	2	&	1	&		&		&		&		&		&		&		&		&		&		&		&		&		&		&		\\
16	&	83	&	68	&	26	&	5	&	2	&	2	&	2	&		&		&		&		&		&		&		&		&		&		&		&		&		&		\\
17	&	175	&	62	&	30	&	12	&	4	&	2	&	2	&	1	&		&		&		&		&		&		&		&		&		&		&		&		&		\\
18	&	407	&	133	&	49	&	21	&	10	&	4	&	2	&	2	&	2	&		&		&		&		&		&		&		&		&		&		&		&		\\
19	&	960	&	285	&	99	&	39	&	18	&	8	&	4	&	2	&	2	&	1	&		&		&		&		&		&		&		&		&		&		&		\\
20	&	2868	&	766	&	179	&	77	&	33	&	15	&	7	&	4	&	2	&	2	&	2	&		&		&		&		&		&		&		&		&		&		\\
21	&	2926	&	847	&	364	&	88	&	43	&	22	&	11	&	6	&	3	&	2	&	2	&	1	&		&		&		&		&		&		&		&		&		\\
22	&	22088	&	5522	&		&	174	&	74	&	36	&	20	&	10	&	6	&	2	&	2	&	2	&	2	&		&		&		&		&		&		&		&		\\
23	&	42968	&	10701	&		&	336	&	126	&	57	&	31	&	16	&	8	&	4	&	2	&	2	&	2	&	1	&		&		&		&		&		&		&		\\
24	&	338016	&	84964	&	80	&	690	&	244	&	102	&	51	&	27	&	14	&	7	&	4	&	2	&	2	&	2	&	2	&		&		&		&		&		&		\\
25	&	649922	&	162986	&	96	&	1402	&	480	&	190	&	83	&	65	&	23	&	12	&	6	&	4	&	2	&	2	&	2	&	1	&		&		&		&		&		\\
26	&	5199376	&	1299844	&	848	&	2974	&	977	&	351	&	148	&	67	&	38	&	20	&	11	&	6	&	4	&	2	&	1	&	2	&	2	&		&		&		&		\\
27	&	10029150	&	2506644	&	848	&	6308	&	1927	&	655	&	262	&	114	&	56	&	32	&	18	&	9	&	5	&	4	&	1	&	1	&	2	&	1	&		&		&		\\
28	&	180226336	&	20056584	&	1536	&	13688	&	3987	&	1310	&	459	&	194	&	93	&	50	&	28	&	15	&	8	&	4	&	4	&	1	&	2	&	2	&	2	&		&		\\
29	&		&	38777664	&	19388832	&	29292	&	8245	&	2599	&	898	&	353	&	155	&	77	&	42	&	24	&	12	&	7	&	4	&	3	&	2	&	2	&	2	&	1	&		\\
30	&	9110544	&	708168	&	77558760	&	61270	&	17677	&	5426	&	1767	&	546	&	266	&	127	&	65	&	36	&	20	&	11	&	7	&	4	&	3	&	2	&	2	&	2	&	2	\\
31	&	150266880	&		&		&		&		&		&		&		&		&		&		&		&		&		&		&		&		&		&		&		&		\\
32	&	300533760	&		&		&		&		&		&		&		&		&		&		&		&		&		&		&		&		&		&		&		&		\\
33	&	583395120	&		&		&		&		&		&		&		&		&		&		&		&		&		&		&		&		&		&		&		&		\\
34	&	1166803110	&		&		&		&		&		&		&		&		&		&		&		&		&		&		&		&		&		&		&		&		\\
35	&	2268771670	&		&		&		&		&		&		&		&		&		&		&		&		&		&		&		&		&		&		&		&		\\
36	&	4537543340	&		&		&		&		&		&		&		&		&		&		&		&		&		&		&		&		&		&		&		&		\\

\hline

\end{tabular}
\caption{Lower bounds on DNA codes satisfying GC and Reverse Complement Constraints for $4 \leq n \leq 36$ and $10 \leq d \leq 30$} 
\label{boundtable2}
\end{sidewaystable}
\end{footnotesize}
\end{center}

\begin{center}
\begin{footnotesize}
\begin{sidewaystable}
\centering

\begin{tabular}{|l|l|l|l|l|l|l|l|l|l|l|l|l|l|l|l|l|l|l|l|}
\hline
n/d	&	2	&	3	&	4	&	5	&	6	&	7	&	8	&	9	&	10	&	11	&	12	&	13	&	14	&	15	&	16	&	17	&	18	&	19	&	20	\\ \hline
4	&	32	&	6	&	2	&		&		&		&		&		&		&		&		&		&		&		&		&		&		&		&		\\
5	&	116	&	32	&	4	&	2	&		&		&		&		&		&		&		&		&		&		&		&		&		&		&		\\
6	&	512	&	62	&	28	&	4	&	2	&		&		&		&		&		&		&		&		&		&		&		&		&		&		\\
7	&	1968	&	196	&	42	&	12	&	2	&	2	&		&		&		&		&		&		&		&		&		&		&		&		&		\\
8	&	8192	&	620	&	128	&	30	&	16	&	2	&	2	&		&		&		&		&		&		&		&		&		&		&		&		\\
9	&	$2887!$	&	1952	&	346	&	80	&	22	&	8	&	2	&	2	&		&		&		&		&		&		&		&		&		&		&		\\
10	&	$2786!$	&	8064	&	2016	&	496	&	120	&	17	&	8	&	2	&	2	&		&		&		&		&		&		&		&		&		&		\\
11	&	$2677!$	&	23565	&	4832	&	607	&	136	&	40	&	15	&	6	&	2	&	2	&		&		&		&		&		&		&		&		&		\\
12	&	$2734!$	&	$2609!$	&	32640	&	4032	&	2016	&	120	&	31	&	12	&	4	&	2	&	2	&		&		&		&		&		&		&		&		\\
13	&		&	65536	&	65280	&	5469	&	2016	&	240	&	70	&	24	&	10	&	4	&	2	&	2	&		&		&		&		&		&		&		\\
14	&	$>=10^7$	&	32640	&	523776	&	32640	&	8192	&	2016	&	512	&	120	&	32	&	8	&	4	&	2	&	2	&		&		&		&		&		&		\\
15	&		&	1047552	&	1047552	&	65280	&	16384	&	4032	&	1024	&	118	&	64	&	17	&	6	&	3	&	2	&	2	&		&		&		&		&		\\
16	&	$>=10^7$	&	$>=10^7$	&	8386560	&		&	130560	&	32640	&	8192	&	480	&	120	&	120	&	32	&	5	&	2	&	2	&	2	&		&		&		&		\\
17	&		&		&	$>=10^7$	&	65280	&	16384	&	65280	&	16384	&	679	&	197	&	68	&	64	&	12	&	4	&	2	&	2	&	2	&		&		&		\\
18	&	$>=10^7$	&	$>=10^7$	&	523776	&	2095104	&	2091504	&		&	16384	&	8064	&	2016	&	143	&	120	&	22	&	10	&	4	&	2	&	2	&	2	&		&		\\
19	&		&	$>=10^7$	&	261888	&		&	1048604	&	131072	&	32512	&	8128	&	1095	&	321	&	109	&	42	&	19	&	8	&	4	&	2	&	2	&	2	&		\\
20	&	$>=10^7$	&	$>=10^7$	&	$>=10^7$	&	523776	&	2095104	&	1046528	&	523776	&	32256	&	1598	&	2016	&	480	&	83	&	35	&	16	&	7	&	4	&	2	&	2	&	2	\\
\hline
\end{tabular}
\caption{Lower bounds on DNA codes satisfying Hamming distance and Reverse Complement Constraints for $4 \leq n \leq 20$ and $2 \leq d \leq 20$}
\label{RCbound}
\end{sidewaystable}
\end{footnotesize}
\end{center}

\section{Future Work and Challenges}
DNA codes designing have received a great deal of attention by researchers in the last decade. In spite of different approaches proposed in the literature for the construction of DNA codes and constraints, it is still a challenge to design the optimal DNA codes.  Classifying the DNA codes for specific application has opportunities to use it for real applications. Though researchers have worked on improvement of the bounds of DNA codes satisfying the set of constraints, designing DNA codes satisfying maximum number of constraints achieving bound is still a huge challenge. Better codes with higher length and distance can be designed by using other algebraic methods or computational methods improving the bounds and obtaining the bounds for the missing $n$ and $d$ can be achieved. Defining DNA code as mathematical structure with the possible operation is interesting area to explore in which different operation can be defined on DNA nucleotides that satisfies the desired properties for computation. There are attempts to develop DNA codes satisfying the thermodynamic constraints though it is a challenge to develop the DNA codes that fits perfectly for the practical application of DNA strands. One of the important research problem is to work on the optimality condition of the DNA codes. To simulate the process of the DNA code designing and practical protocols involved in the DNA computation, it can be automated by developing a platform where DNA codes can be designed and simulated to check with the performance and accuracy. With emerging area of algebraic coding and biological coding theory, these challenges can be investigated and resolved.
\bibliographystyle{IEEEtran} 
\bibliography{dnacodesreferences}

\begin{appendices}

\section{Bounds on DNA Codes} \label{App:AppendixA}

\begin{enumerate}
\item\textit{Johnson type Bound-} This bound is derived by shortening the code to length $n-1$. The code is modified by choosing the codewords with a fix character $b \in Z_q$ at the $i_{th}$  position where $i \in \{1 \ldots n\} $ and deleting $i_{th}$  position from the codewords. \\ For $0 \leq d \leq n$ and $0 < w < n$, 
\begin{enumerate}
\item ${A}_{4}^{GC}(n,d,w) \leq \lfloor \frac{2n}{w} {A}_{4}^{GC}(n-1,d,w-1)\rfloor$ \cite{king2003bounds}.\newline
\item ${A}_{4}^{GC}(n,d,w) \leq \lfloor \frac{2n}{n-w} {A}_{4}^{GC}(n-1,d,w)\rfloor$\cite{king2003bounds}. \newline
\item ${A}_{{4}}^{{R}}(n,d) \leq \lfloor \frac{1}{4} {A}_{{4}}^{{R}}(n-1,d) \rfloor$ \cite{ignatova2008dna}.
\end{enumerate}

\item\textit\textbf{Halving Bounds-} 
\begin{enumerate}
\item This is motivated by the fact DNA code $\mathscr{C}_{DNA}$ and reverse DNA code  $\mathscr{C}_{DNA}^\textbf{R}$ are disjoint.
Let $n \geq 1$ be an integer. For each integer d with $0 < d\leq n$, 
${A}_{{4}}^{{R}}(n,d) \leq \dfrac{1}{2} {A}_{{4}}(n,d)$ \cite{ignatova2008dna}. \newline
\item
For $0 < d \leq n$ and $0\leq w\leq n$, ${A}_{{4}}^{{GC,RC}}(n,d,w) \leq \frac{1}{2}{A}_{{4}}^{{GC}}(n,d,w)$\cite{ignatova2008dna}.
\item
For $0 < d \leq n$ and $0\leq w\leq n$,${A}_{{4}}^{{GC,R}}(n,d,w) \leq \frac{1}{2}{A}_{{4}}^{{GC}}(n,d,w)$\cite{ignatova2008dna}.
\newline
\end{enumerate}

\item\textit{Gilbert-Type Bounds-}${A}_{{4}}^{{GC}}(n,d,w)$ is derived by dividing the total number of DNA strings of length $n$ with $GC$-content $w$ by the number of DNA string with distance $d-1$ from the fixed codeword. 
\begin{enumerate}
\item   For $0 \leq d \leq n$ and $0\leq w\leq n$,
\newline${A}_{{4}}^{{GC}}(n,d,w) \geq \dfrac{\binom{n}{w}2^{w}2^{n-w}}{\sum_{r=0}^{d-1}\sum_{i=0}^{min{\lfloor r/2 \rfloor,w,n-w}} \binom{w}{i}\binom{n-w}{i} \binom{n-2i}{r-2i} 2^{2i}}$ \cite{ignatova2008dna}.

\item 
For $0 \leq d \leq n$ and $0\leq w\leq n$,
\newline${A}_{{4}}^{{GC}}(n,d,w) \geq \dfrac{\binom{n}{w}2^{n}}{\sum_{r=0}^{d-1}\sum_{i=0}^{min{\lfloor r/2 \rfloor,w,n-w}} \binom{w}{i}\binom{n-w}{i} \binom{n-2i}{r-2i} 2^{2i}}$ \cite{king2003bounds}.
\end{enumerate} 

\item This bound is based on the aspect of $GC$-content of given DNA codeword is equal to reverse of DNA codeword \cite{king2003bounds}. \\ For $0 \leq d \leq n$ and $0 \leq w \leq n$,\newline 
	\begin{enumerate}
	\item ${A}_{{4}}^{{GC,RC}}(n,d,w) = {A}_{{4}}^{{GC,R}}(n,d,w)$ if n is even.
	\newline
	\item
	${A}_{{4}}^{{GC,RC}}(n,d,w) \leq {A}_{{4}}^{{GC,R}}(n+1,d+1,w)$ if n is odd.
	\\
	\item ${A}_{{4}}^{{GC,R}}(n,d+1,w) \leq {A}_{{4}}^{{GC,R}}(n,d,w) \leq$  ${A}_{{4}}^{{GC,R}}(n,d-1,w) $	if n is odd.\\
	\end{enumerate}
	
\item  By using ${A}_{{4}}^{{GC}}(n,d,w) \geq {A}_{{2}}(n,d,w) \cdot {A}_{{2}}(n,d)$ inequality, following bound is derived\cite{king2003bounds}. \\ For $0 \leq w \leq n$, ${A}_{{4}}^{{GC}}(n,2,w) = \dbinom{n}{w} 2^{n-1}$.
\newline

\item By using Halving bound ${A}_{{4}}^{{GC,RC}}(n,d,w) \leq \frac{1}{2}{A}_{{4}}^{{GC}}(n,d,w)$ for $d=2$ one can calculate the bound \cite{king2003bounds}.\\
For $0 \leq w \leq n$ and n is even,\newline \\ ${A}_{{4}}^{{GC,RC}}(n,2,w) = \dbinom{n}{w} 2^{n-2}.$
\newline

\item Bound ${A}_{{4}}^{{GC}}(n,d,w)$ is computed by dividing the total number of words with $GC$-content $w$ that are at distance at least $d$ from their reverse-complements by the number of these codewords that are at distance at most $d − 1$ from any fixed
codeword  \cite{king2003bounds}.\\

For $0 \leq d \leq n$ and $0\leq w\leq n$,
\newline \newline${A}_{{4}}^{{GC}}(n,d,w) \geq \dfrac{\sum_{r=d}^{n} V(n,d,r)}{2 \sum_{r=0}^{d-1}\sum_{i=0}^{min{\lfloor r/2 \rfloor,w,n-w}} \binom{w}{i}\binom{n-w}{i} \binom{n-2i}{r-2i} 2^{2i}}$

\item \textit{Product Bounds -} This is based on the construction of the DNA code ${A}_{{4}}^{{GC}}(n,d,w)$ with length $n$, minimum Hamming distance $d$ and $GC$-content $w$ from binary constant-weight codes ${A}_{{2}}(n,d,w)$ and ternary constant-weight codes ${A}_{{3}}(n,d,w)$ with length $n$, Hamming weight $w$ and minimum Hamming distance $d$ \cite{king2003bounds} \cite{ignatova2008dna}. 

For $0 \leq d \leq n$ and $0 \leq w \leq n$,\newline 
	\begin{enumerate}
	\item ${A}_{{4}}^{{GC}}(n,d,w) \geq {A}_{{2}}(n,d,w) \cdot {A}_{{2}}(n,d)$ \newline
	\item ${A}_{{4}}^{{GC,R}}(n,d,w) \geq {A}_{{2}}^{{R}}(n,d,w) \cdot {A}_{{2}}(n,d)$ \newline
	\item ${A}_{{4}}^{{GC,R}}(n,d,w) \geq {A}_{{2}}(n,d,w) \cdot {A}_{{2}}^{{R}}(n,d)$ \newline
	\item ${A}_{{4}}^{{GC}}(n,d,w) \geq {A}_{{3}}(n,d,w) \cdot {A}_{{2}}(n-w,d)$ \newline
	\item ${A}_{{4}}^{{GC,R}}(n,d,w) \geq {A}_{{3}}^{{R}}(n,d,w) \cdot {A}_{{2}}(n-w,d)$ \newline
	\item ${A}_{{4}}^{{GC,R}}(n,d,w) \geq {A}_{{3}}(n,d,w) \cdot {A}_{{2}}^{{R}}(n-w,d)$ \newline	
	\item ${A}_{{4}}^{{R}}(n,d) \geq {A}_{{2}}^{{R}}(n,d) \cdot {A}_{{2}}(n,d)$.\newline
	\end{enumerate}

\item \textit{Reverse and Reverse complement codes bounds -} This can be simple observed by the reverse and reverse complement property of the DNA codeword \cite{ignatova2008dna} \cite{marathe2001combinatorial}. \\
Let $n \geq 1$ be an integer,
	\begin{enumerate}
	\item If n is even, then ${A}_{{4}}^{{RC}}(n,d) = {A}_{{4}}^{{R}}(n,d)$ and \newline
	\item If n is odd, then ${A}_{{4}}^{{RC}}(n,d) \leq {A}_{{4}}^{{R}}(n+1,d+1)$
	\end{enumerate}

\item\textit{Special cases - } Bounds are observed by considering different combination of length $n$ and $GC$-content $w$ \cite{marathe2001combinatorial}.

For $n > 0 $, with $0 \leq d \leq n$ and $0\leq w\leq n$,
\begin{enumerate}
\item ${A}_{{4}}^{{GC}}(n,d,0) = {A}_{{2}}(n,d)$ \newline
\item ${A}_{{4}}^{{GC}}(n,d,w) = {A}_{{4}}^{{GC}}(n,d,n-w)$ \newline
\item ${A}_{{4}}^{{GC}}(n,n,w) = $  
\[ \left \{
  \begin{tabular}{c}
  $4$ if $w = n/2$ \\
  $3$ if $n \leq w < n/2$ or $n/2 < w \leq 2n/3$  \\
  $2$ if $w < n/3$ or $w > 2n/3$  \\
  \end{tabular}
  \right \}
\]\newline

\item ${A}_{{4}}^{{GC,RC}}(n,n,w) = $\[ \left \{\begin{tabular}{c} $2$ if $w = n/2$ \\ $1$ if $w\neq n/2$ \\ \end{tabular}
  \right \}
\]
\item ${A}_{{4}}^{{GC}}(n,1,w) = \dbinom{n}{w}2^{n} $

\end{enumerate}

\item \textit{Bounds on reverse code for $d=3$ -} This bound is computed from the concept of sphere-packing bound for codes \cite{marathe2001combinatorial}. \\ ${A}_{q}^{R}(n,3) \leq \dfrac{q^{\lceil n/2 \rceil} \sum {i=2} {\lfloor n/2 \rfloor} \binom {\lfloor n/2 \rfloor}{i} (q-1)^{i}}{2(1+4(q-2)+(n-4)(q-1))}$.
\\ 
\item \textit{Bounds on reverse code of size $S$ -} By using Greedy approach to calculate size of the code, bound is derived \cite{marathe2001combinatorial}. \\

Let $V(s,d)$ be the number of words of $S$ such that they have distance $d$ from s where $s \varepsilon S$.
\begin{enumerate}
\item ${A}_{q}^{R}(n,d) \leq \dfrac{\vert S \vert}{2V^{-}(\lfloor (d-1)/2 \rfloor)}$ where
$V^{-} (d) = min\{V(s,d) \vert s \varepsilon S\}$.
\\
\item ${A}_{q}^{R}(n,d) \geq \dfrac{\vert S \vert}{2V^{+} d-1}$ where,
$V^{+}(d) = max\{V(s,d)\vert s \varepsilon S\}$.
\end{enumerate}

\item\textit{Bounds on reverse code for $d=2$ - } Bounds on reverse code for $d=2$ is obtained by claims 
\begin{itemize}
\item Any two words from the same subset differ in at least two positions ie. $d_{H}(\textbf{x}_{\textbf{DNA}}$, $\textbf{y}_{\textbf{DNA}})$ $ \geq 2 $ $ \forall $ $\textbf{x}_{\textbf{DNA}}$, $\textbf{y}_{\textbf{DNA}} \in  \mathscr{C}_{DNA}$.
\item If a word belongs to a subset, its reversal is also in the same subset ie. if $\textbf{x}_{\textbf{DNA}} \in \mathscr{C}_{DNA}$ then $\textbf{x}_{\textbf{DNA}}^{\textbf{R}} \in \mathscr{C}_{DNA}$
\item All the $q^{n/2}$ palindromes are in the same subset.

\end{itemize}

\begin{enumerate}
\item ${A}_{q}^{R}(n,2) = \dfrac{q^{n-1}}{2}$, for even $n$ and $q\varepsilon\{2,4\}$, and \\
\item ${A}_{q}^{R}(n,2) = \dfrac{q^{n-1} - q^{\lfloor n/2 \rfloor}}{2}$, for odd $n$ and $q\varepsilon\{2,4\}$
\end{enumerate}
\item\textit{Doubling Construction - } This is motivated by the minimum Hamming distance between the DNA codeword, its reverse codeword and revere complement. It is observed from the property of DNA code with $d_{H}(\textbf{x}_{\textbf{DNA}}$, $\textbf{y}_{\textbf{DNA}})$ $\geq d $, $H_{DNA}(\textbf{x}_{\textbf{DNA}}^{\textbf{R}},\textbf{y}_{\textbf{DNA}}) \geq d$, $H_{DNA} (\textbf{x}_{\textbf{DNA}},\textbf{y}_{\textbf{DNA}}^{\textbf{C}}) \geq d $ and $H_{DNA}(\textbf{x}_{\textbf{DNA}},\textbf{y}_{\textbf{DNA}}^{\textbf{RC}}) \geq d $ \\ 

For $n\geq 2$, ${A}_{2}^{R}(2^{n},2^{n-1})=2^{n}$\cite{marathe2001combinatorial}.
\vspace{0.1cm}
\item 
\begin{enumerate}

\item \textit{Bounds on even and odd length $n$ reverse code -} In this bounds, maximum size of reverse code of even and odd length $n$ and relationship between reverse code of even and odd length $n-1$ is demonstrated \cite{marathe2001combinatorial}. \\ \\  ${A}_{q}^{R}(n-1,d) \leq {A}_{q}^{R}(n,d) \leq {A}_{q}^{R}(n,d-1)$ and \\
\item ${A}_{q}^{R}(n-1,d)\geq {A}_{q}^{R}(n,d)/q$, for odd $n$.
\end{enumerate}

\item \textit{Bounds on Hamming distance constraint-} V.Phan provided lower and upper bounds of DNA codeword sets which satisfy the h-distance constraint \cite{zhang2011bounds}.
\center $\dfrac{4^{n-d+1}}{d \binom {n}{d-1}} \leq \vert S \vert \leq 4^{n}$

\end{enumerate}
\end{appendices}
\end{document}